\documentclass{jpsj3}
\usepackage{txfonts}
\usepackage{color}

\title
{Vanishing of Resistivity upon Freezing of Vortex Liquid in Clean 
Superconductors}

\author{Naratip Nunchot and Ryusuke Ikeda}

\inst{Department of Physics, Graduate School of Science, Kyoto University, Kyoto 606-8502, Japan} 

\abst{Superconducting transition, defined as vanishing of the resistivity, under a magnetic field in a clean bulk type II superconductor with weak sample disorder is believed to be a reflection of freezing of the vortex liquid to a kind of vortex solids. This fundamental issue on superconductivity is examined in detail. Based on the Ginzburg-Landau fluctuation theory for a three-dimensional (3D) system and through a supplementary study in 2D case, we find that the resistivity in the weakly disordered 3D case vanishes in a nearly discontinuous way, reflecting growth of the Bragg peaks on approaching the vortex lattice melting transition. In contrast, such a sharp decrease of the resistivity does not clearly appear in the corresponding 2D case. The consequences of this difference in the vanishing behavior of the resistivity between the 2D and 3D systems are discussed in relation to available experimental facts.}

\begin{document}

\maketitle

\section{Introduction}
When a three-dimensional (3D) bulk type II superconductor is perfectly clean, the resistivity $\rho_\perp$ in a finite magnetic field and under a current perpendicular to the field remains finite even in the low temperature limit because of the vortex flow \cite{dG}. Therefore, a small but nonvanishing amount of disorder is inevitably assumed to be present in real clean superconducting (SC) materials showing vanishing of $\rho_\perp$ at a finite temperature $T_\rho$. A nearly discontinuous vanishing of $\rho_\perp$, observed in high quality samples of such materials under much lower magnetic fields than the zero temperature depairing field $H_{c2}(0)$, is regarded as a consequence of a glass transition \cite{Natter,Blatter}. As the ground state of the field-induced vortices in lower fields, the Bragg-glass or the elastic glass was proposed \cite{Natter,GleD}. It is believed that, on cooling to enter this glass phase, the resistivity will vanish discontinuously reflecting the first order freezing transition from the vortex liquid to the vortex solid in the perfectly clean limit. However, the realization of a vortex glass (VG) in moderately clean cases, where large bundles of positionally correlated vortices are formed, has also been proposed earlier \cite{Blatter,FFH}. In this picture, the vanishing of $\rho_\perp$ does not have to discontinuously occur. Experimentally, situations \cite{Worthington,Nishizaki} in which a nearly discontinuous decrease of $\rho_\perp$ does not directly reach complete superconductivity have been seen repeatedly in clean samples. Therefore, an established theory on the nearly discontinuous vanishing of $\rho_\perp$ in moderately clean 3D type II superconductors still remains unavailable. 

Similarly, a confirmed picture remains unavailable on the temperature dependences of $\rho_\perp$ in relatively clean 2D systems of field-induced point 
vortices. In such 2D systems, a nearly discontinuous vanishing of $\rho_\perp$ is rarely seen even in relatively clean materials \cite{Saito}. It is unclear at present whether this fact implies that the 2D vortex lattice melting in clean limit consists of two-step continuous transitions \cite{HN}. 

In this work, the effects of freezing of the 2D and 3D vortex liquids on $\rho_\perp$ are studied in detail based on the high field Ginzburg-Landau (GL) fluctuation theory for a superconductor with weak point-like disorder included. It is well known that, in high fields, the SC fluctuation contributing to thermodynamics is dominated by its lowest Landau level (LLL) modes \cite{Thouless,BNT,IOT89,IOT91,Moore89,Dorsey}. On the Gaussian level, the VG fluctuation formulated in a form including effects of the vortex solidification was discussed \cite{RI96} in the case with weak enough disorder, and a sudden enhancement of the VG fluctuation accompanying the first order vortex lattice melting in clean limit was argued to be the origin of the nearly discontinuous vanishing of $\rho_\perp$ observed in high quality samples of high $T_c$ cuprates \cite{Kwok}. However, no calculation results on $\rho_\perp$ based on a detailed analysis including the interplay between the disorder and the mutual interaction among the SC fluctuations have been presented previously. 

In the present analysis, the parquet resummation technique \cite{YM,Tsuneto} is used to evaluate the interplay between the fluctuation-interaction and the disorder and to compute the renormalized vertex correction to the impurity strength which forms the backbone of the VG correlation function. This treatment can be fully performed in 2D, while the corresponding 3D results will be approximately derived by utilizing the 2D results. It is shown that, in 3D case, the resistivity shows a clear sudden drop, reflecting a sharp growth of the Bragg peak of the vortex structural factor upon cooling \cite{RI96}, although the assumed VG transition is a {\it continuous} one. In contrast, such a sharp drop of the resistivity is not clearly seen in 2D case: There, it appears only faintly close to the end of the vanishing resistivity curves. Both in 2D and 3D cases with quite weak disorder, however, the temperature at which the resistivity shows a sharp reduction is in good agreement with the vortex lattice melting temperature estimated previously \cite{HT,RI95} based on experimental results and the numerical simulation in clean limit performed in the LLL approximation \cite{KN,Saiki}. These calculation results will be discussed through comparison with available experimental results \cite{com}. 

The present paper is organized as follows. In sec.2, we introduce the GL model and explain how the mass renormalization is performed in the model by focusing on the 2D case. The parquet analysis in 2D case is explained in sec.3, and the content in sec.2 and 3 is extended to the 3D case in sec.4. The VG fluctuation in clean enough systems is discussed in sec.5, and the conductivity arising from the VG fluctuation is calculated in sec.6. In sec.7, theoretical pictures following from numerical results on the resistivity are presented, and comments relevant to experimental data are given in sec.8. Some details on theoretical analysis are given in 
Appendices A and B. 

\section{Model}

The original model in studying a superconducting transition is conventionally the  Ginzburg-Landau (GL) Hamiltonian for the order parameter field $\Delta({\bf  r})$. In the case of a thin film with thickness $d$ shorter than the zero temperature coherence length $\xi_0$, i.e., in 2D case, 
the GL Hamiltonian is given by 
\begin{equation}
{\cal H}_{\rm GL} = d N(0) \int d^2{\bf r} \biggl[ (t-1) |\Delta|^2 + u({\bf r}) |\Delta|^2 + \frac{g_0}{2} |\Delta|^4 + \xi_0^2 \biggl|\biggl( - i \nabla + \frac{2 \pi}{\phi_0} {\bf A} \biggr) \Delta \biggr|^2 \biggr], 
\label{2dgl}
\end{equation}
where $t=T/T_{c0}$ is the normalized temperature expressed by the zero field SC transition temperature $T_{c0}$, $\phi_0$ is the flux quantum, and $N(0)$ is the density of states of the normal electron system. Further, a random potential $u({\bf r})$ expressing a spatial variation of $T_{c0}$ has been introduced for later convenience. 
By noting that the Boltzmann factor is $\exp(-{\cal H}_{\rm GL}/k_{\rm B} T)$,  however, we will choose to start from the normalized form 
\begin{equation}
{\cal H}[\psi] = {\cal H}_{\rm GL}/(k_{\rm B} T) = \xi_0^{-2} \!\! \int d^2r \, \biggl[ (t-1)\rho({\bf r}) + u({\bf  r})\rho({\bf  r}) + \pi t b^{(2)}_G |\rho({\bf  r})|^2 
+ \xi_0^2 \, \biggl| \biggl( -i \nabla+\frac{2\pi}{\phi_0}{\bf A}({\bf r}) \biggr) \psi({\bf r}) \biggr|^2  \biggr]
\label{r2dgl}
\end{equation}
where $\psi = \Delta \, \sqrt{d \xi_0^2 N(0)/k_{\rm B} T}$, $\rho=|\psi|^2$, and $b_{\rm G}^{(2)} = g_0 k_{\rm B} T_{c0}/(2 \pi \xi_0^2 d N(0))$. The randomness with respect to $u({\bf r})$ is defined by $\overline{u({\bf r})} = 0$ and 
\begin{equation} 
\overline{u({\bf r}) u({\bf r}')} = \xi_0^2 \Delta^{(2)} \delta^{(2)}({\bf r}-{\bf r}'), 
\label{impcor}
\end{equation}
where the overbar implies that the random average was taken. 
Hereafter, all the length scales will be normalized by $\xi_0$. 

Consistently with the Abrikosov mean field theory \cite{AAA}, the order parameter will be expanded in terms of the ortho-normalized LLL eigenfunctions $U_p({\bf r})$. In the Landau gauge 
\begin{equation}
{\bf A} = H x {\bf e}_y 
\label{Lgauge}
\end{equation}
with the unit vector ${\bf e}_y$ in the $y$-direction, 
they are expressed as 
\begin{eqnarray}
\psi({\bf r}) &=& \sum_p \varphi_p U_p({\bf r}), \nonumber \\
U_p({\bf r}) &=& \biggl( \frac{h}{\pi {\tilde L}^2} \biggr)^{1/4} \exp\biggl(ipy - \frac{h}{2} \biggl(x + \frac{p}{h} \biggr)^2 \biggr), 
\end{eqnarray}
where $\varphi_p$ is the LLL fluctuation field with the index $p$ expressing the degree of freedom of the guiding center and counting the degeneracy in LLL, ${\tilde L}$ is the system size normalized by $\xi_0$, $2 \pi r_H^2 = \phi_0/H$ is the area per vortex, and $h=H/H_{c2}(0)= 2 \pi \xi_0^2 H/\phi_0$ is the dimensionless strength of the magnetic field. Here, the number $N_v$ of degeneracy in LLL is given by 
\begin{equation}
N_v = \sum_p = \frac{HS}{\phi_0} = \frac{{\tilde S} h}{2 \pi},  
\end{equation}
where ${\tilde S}$ is the system area divided by $\xi_0^2$. 
Then, using the Fourier transform of $\rho$, i.e., 
\begin{equation}
\rho({\bf k}) = \frac{1}{{\tilde S}^{1/2}} \sum_{p} {} \exp(i p k_1/h) \, \varphi^*_{p-\frac{k_2}{2}} \, \varphi_{p+\frac{k_2}{2}}, 
\end{equation}
${\cal H}[\psi]$ becomes 
\begin{equation}
{\cal H}[\psi] = \sum_p \, \mu_0 |\varphi_p|^2 + t \pi b_{\rm G}^{(2)} \sum_{\bf k} v_{\bf k} \, \rho({\bf k}) \rho(-{\bf k}) + \sum_{\bf k} u_{-{\bf k}} v_{\bf k}^{1/2} \rho({\bf k}), 
\label{LLL2dgl}
\end{equation}
where ${\bf k}=(k_1, k_2)$, $\mu_0 = -1+t+h$ is the bare mass, i.e., the mass of the unrenormalized 
LLL fluctuation, and 
\begin{equation}
v_{\bf k} = \exp\biggl(- \frac{{\bf k}^2}{2h} \biggr). 
\end{equation}
Further, eq.(\ref{impcor}) is rewritten as 
\begin{equation}
\overline{u_{\bf k} u_{-{\bf k}'}} = \Delta^{(2)} \delta_{k_1, k_1'} \delta_{k_2, k_2'}. 
\end{equation}

Since, more or less, the disordered phase of a random system is treated in this work, the replica trick \cite{EA} will be introduced for the free energy $F$ in the 
manner 
\begin{equation}
- \frac{F}{k_{\rm B} T} = \overline{{\rm log}Z} = \lim_{n \to +0} \frac{\overline{Z^n} - 1}{n}, 
\end{equation}
where 
\begin{eqnarray}
\overline{Z^n} &=& \int \prod_{a=1}^n {\cal D}(\varphi^{* a}, \varphi^a) \, \exp(-S_0 - S_I), \nonumber \\
S_0 &=& \sum_{a=1}^n \sum_p \mu_0 \, |\varphi_p^a|^2, \nonumber \\
S_I &=& \! \frac{t \pi b^{(2)}_G}{2 {\tilde S}} \! \! \sum_{a,b,c,d=1}^n \sum_{p, p', {\bf k}} \exp(i (p-p')k_1/h) [\delta_{a,c} \delta_{b,d} h_{bare, ab}({\bf k}) \nonumber \\
 &+& \delta_{a,d} \delta_{b,c} {\hat h}_{bare, ab}({\bf k}) ] \varphi^{* a}_{p_-} \varphi^{* b}_{p'_+} \varphi^{c}_{p_+} \varphi^{d}_{p'_-}, 
\end{eqnarray}
where the indices $a,b,c,$ and $d$ denote the replica indices, $p_\pm = p \pm k_2/2$, $p'_\pm = p' \pm k_2/2$, 
\begin{eqnarray}
h_{bare, ab}({\bf k}) &=& v_{\bf k} \, 
(\delta_{a,b} - \theta(t)), \nonumber \\ 
\theta(t) &=& \frac{\Delta^{(2)}}{2 \pi t b_G^{(2)}},  
\end{eqnarray}
and 
\begin{equation}
{\hat h}_{bare, ab}({\bf k}) = \frac{1}{N_v} \sum_{{\bf k}'} \exp(i ({\bf k}' \times {\bf k})_z/h) \, h_{bare, ab}({\bf k}').  
\end{equation}
Here, for later convenience, ${\hat h}_{bare, ab}$ ($=h_{bare, ab}$) was introduced to clarify the symmetry between the two "particle-hole" channels. 
In fact, it is found by examining \cite{IOT90} the perturbation series for the four-point vertex function that this symmetric form is preserved in any order of the perturbation series. That is, in the corresponding form to the bare interaction action $S_I$, the fully renormalized interaction action takes the form  
\begin{equation}
S_{R,I} = \frac{t \pi b^{(2)}_G}{\tilde S} \sum_{a,b,c,d=1}^n \sum_{p, p', {\bf k}} \exp(i (p-p')k_1/h) \, \Gamma_{ab,cd}({\bf k}) \, \varphi^{* a}_{p_-} \varphi^{* b}_{p'_+} \varphi^{c}_{p_+} \varphi^{d}_{p'_-}, 
\end{equation}
where
\begin{eqnarray}
\Gamma_{ab,cd}({\bf k}) &=& \frac{1}{2} [
\delta_{a,c} \delta_{b,d} h_{R, ab}({\bf k}) + \delta_{a,d} \delta_{b,c} {\hat h}_{R, ab}({\bf k})], \nonumber \\
{\hat h}_{R, ab}({\bf k}) &=& \frac{1}{N_v} \sum_{{\bf k}'} \exp(i ({\bf k}' \times {\bf k})_z/h) \, h_{R, ab}({\bf k}'),
\end{eqnarray}
and $h_{R,ab}({\bf k})$ is a quantity to be derived self-consistently and takes the form 
\begin{equation}
h_{R,ab}({\bf k}) = \delta_{a,b} f({\bf k}) - \theta(t) \, w({\bf k}). 
\end{equation}
We note that $f({\bf k})$ and $w({\bf k})$ are also dependent on $\theta(t)$. 

Next, we explain how the physical quantities directly reflecting the positional correlations between the vortices are defined in the present replica formalism. First, let us consider the four-body correlation function $\langle \varphi_1^* \varphi_2^* \varphi_3 \varphi_4 \rangle$ which can be defined in clean limit with no randomness. To directly see the positional correlation between the vortices, it is often convenient to focus on the connected part of this correlation function in the manner 
\begin{equation}
\langle \varphi_1^* \varphi_2^* \varphi_3 \varphi_4 \rangle_c \equiv \langle \varphi_1^* \varphi_2^* \varphi_3 \varphi_4 \rangle - \langle \varphi_1^* \varphi_3 \rangle \langle \varphi_2^* \varphi_4 \rangle - \langle \varphi_1^* \varphi_4 \rangle \langle \varphi_2^* \varphi_3 
\rangle.
\end{equation}
On the other hand, this quantity is replica-diagonal in the replica formalism and is expressed in the form 
\begin{equation}
\overline{\langle \varphi_1^* \varphi_2^* \varphi_3 \varphi_4 \rangle} = \lim_{n \to +0} \frac{1}{n} \sum_{a=1}^n \langle \varphi_1^{a *} \varphi_2^{a *} \varphi_3^a \varphi_4^a \rangle. 
\end{equation}
Further, in the case of focusing on the fully renormalized vertex function $\Gamma({\bf k})$, it is convenient to use the relation between the connected part of the correlation function and $\Gamma({\bf k})$
\begin{equation}
\overline{\langle \varphi_{p_-}^* \varphi_{p'_+}^* \varphi_{p_+} \varphi_{p'_-} \rangle_c} = -  \frac{4 \pi t b_G^{(2)}}{{\tilde S} \mu^4} \sum_{k_1} \, \exp(i(p-p')k_2/h) \, \Gamma({\bf k}), 
\end{equation}
where $\mu = 1/\langle |\varphi_p|^2 \rangle$, and $\Gamma({\bf k}) = \Gamma_{aa,aa}({\bf k})$. 

\subsection{Structural Factor}

The quantity measuring the vortex positional order is the two-body correlation function on the density $\rho = |\psi|^2$ defined in the manner 
\begin{equation}
f({\bf R}) \equiv \frac{\overline{[\langle (\rho({\bf r}) - \rho_0)(\rho({\bf r}+{\bf R}) - \rho_0) \rangle]_{\bf r}}}{\rho_0^2} = \rho_0^{-2} \overline{[\langle \rho({\bf r}) \rho({\bf r}+{\bf R}) \rangle ]_{\bf r}} - 1, 
\end{equation}
where $\rho_0 = \overline{\langle \rho({\bf k}=0) \rangle}$, and, in LLL, it takes the form
\begin{equation}
f({\bf R}) = \rho_0^{-2} \sum_{\bf k} e^{i{\bf k}\cdot{\bf R}} \, v_{\bf k} \, \overline{\langle \rho({\bf k}) \rho(-{\bf k}) \rangle} - 1.
\end{equation}
Further, using eqs.(7), (18), and (20), we find 
\begin{eqnarray}
\overline{\langle \rho({\bf k}) \rho(-{\bf k}) \rangle} &=& {\tilde S} \sum_{p,p'} \, \exp(i (p-p')k_1/h) \, \overline{\langle \varphi_{p_-}^* \varphi_{p'_+}^* \varphi_{p_+} \varphi_{p'_-} \rangle} \nonumber \\
&=& \rho_0^2 [\delta_{k_1,0} \delta_{k_2,0} + 1 - 2x \Gamma({\bf k})], 
\end{eqnarray}
where $\rho_0 = {\tilde S}^{-1/2} N_v \mu^{-1}$, and 
\begin{equation}
x=\frac{t \, h \, b_G^{(2)}}{\mu^2}. 
\end{equation}

In this way, we choose 
\begin{equation}
f({\bf k}) = 1 - 2x \Gamma({\bf k})
\label{vcorigin}
\end{equation}
as the quantity corresponding to the structural factor, because the original correlation function is given by 
\begin{equation}
f({\bf R}) = \sum_{\bf k} e^{i{\bf k}\cdot{\bf R}} v_{\bf k} f({\bf k}).
\end{equation}

\begin{figure}
\includegraphics{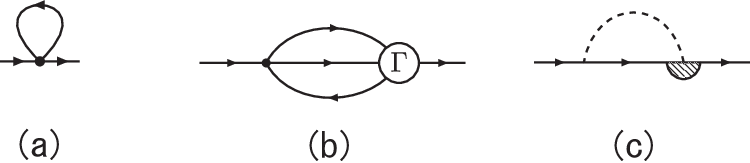}
\caption{Diagrams expressing the self-energy of the SC fluctuation in LLL. Each line with an arrow indicates a LLL fluctuation propagator, the solid dot denotes the bare interaction vertex, the open circle indicated as $\Gamma$ denotes the fully renormalized vertex, and the dashed curve is the pair-wised random potential occurring after the random average. The hatched semicircle is a vertex correction to the bare random potential, and its definition is explained by Fig.2. }
\label{fig.1}
\end{figure}

\begin{figure}
\includegraphics{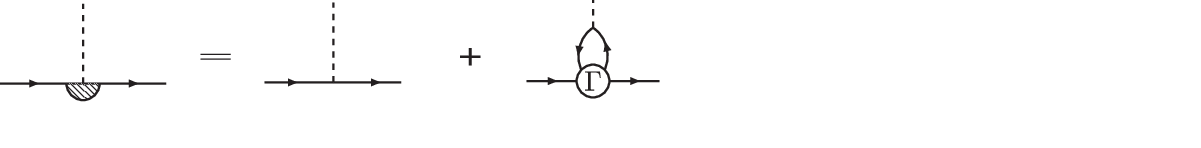}
\caption{Details of the hatched semicircle which includes the vertex correction to the bare random potential. }
\label{fig.2}
\end{figure}

\subsection{Renormalized Mass}

Next, the diagrams expressing the mass renormalization will be examined. Since these diagrams shown in Figs.1 and 2 are diagonal in the replica indices, their expressions may be written in terms 
only of the notation used for the clean limit. 

As usual, the mass renormalization is expressed in terms of the self-energy terms $\Sigma_1^{(2)}+\Sigma_2^{(2)}$ in the form 
\begin{equation}
\mu= t+h-1 + \Sigma_1^{(2)} + \Sigma_2^{(2)}. 
\end{equation}
The self-energy contribution $\Sigma_1^{(2)}$ appearing even in clean limit is given by 
\begin{equation}
\Sigma_1^{(2)} = 2 \mu x [ \, 1 - N_v^{-1} x \sum_{\bf k} v_{\bf k} \Gamma({\bf k}) \, ]. 
\end{equation}
The first term corresponds to Fig.1 (a) which is the Hartree-Fock term. The vertex correction (shaded hemisphere region) in Fig.1 (c) is given by Fig.2 and the vertex $\Lambda(p,p')$ expressing Fig.2 is 
\begin{eqnarray}
\Lambda(p,p') &=& \langle \varphi_p \varphi_{p'}^* \sum_{\bf k} u_{-{\bf k}} v_{\bf k}^{1/2} \rho({\bf k}) \rangle 
- \pi t b_G^{(2)} \langle \varphi_p \varphi_{p'}^* \sum_{{\bf k}'} \Gamma({\bf k}') \rho({\bf k}') \rho(-{\bf k}') \sum_{\bf k} u_{-{\bf k}} v_{\bf k}^{1/2} \rho({\bf k}) \rangle \nonumber \\
&=& {\tilde S}^{-1/2} \mu^{-2} \sum_{\bf k} \delta_{p', p+k_2} u_{-{\bf k}} \, v_{\bf k}^{1/2} \, \exp(i k_1(p+p')/2h) \nonumber \\
&\times& \biggl[ 1 - \frac{2 x}{N_v} \sum_{{\bf k}'} \Gamma({\bf k}') \exp(i(k_1'(p'-p) - k_2' k_1)/h) \biggr]. 
\label{lambdazero}
\end{eqnarray}
Using this expression, $\Sigma_2^{(2)}$ corresponding to Fig.1 (c) is simply expressed as 
\begin{eqnarray}
\Sigma_2^{(2)} &=& - \frac{\mu x \theta(t)}{N_v} \sum_{\bf k} v_{\bf k} \biggl( 1 - \frac{2x}{N_v} \sum_{{\bf k}'} \Gamma({\bf k}') \exp(i({\bf k}' \times {\bf k})_z/h) \biggr) \nonumber \\
&=& - \, \frac{\mu x \theta(t)}{N_v} \sum_{\bf k} v_{\bf k} ( 1 - 2x \Gamma({\bf k}) ). 
\end{eqnarray}
Using the fact that, since the bare interaction vertex $v_{\bf k}$ is isotropic in ${\bf k}$ and a function of ${\bf k}^2$, $\Gamma({\bf k})$ is also a function of ${\bf k}^2$ as far as no transition occurs, the mass renormalization is finally defined by the self-consistent relation 
\begin{equation}
\mu = t-1+h + (2-\theta)\mu x - \frac{1-\theta}{h} \mu x^2 \int_0^\infty d(p^2) v_{\bf p} \Gamma({\bf p}). 
\label{rmass2}
\end{equation} 

\begin{figure}
\includegraphics{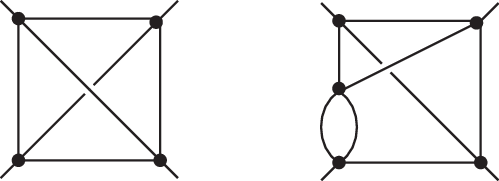}
\caption{Examples of the nonparquet diagrams. The solid dots represent the bare vertices. For simplicity, the arrows of the fluctuation propagators are not drawn. }
\label{fig.3}
\end{figure}

\begin{figure}
\includegraphics{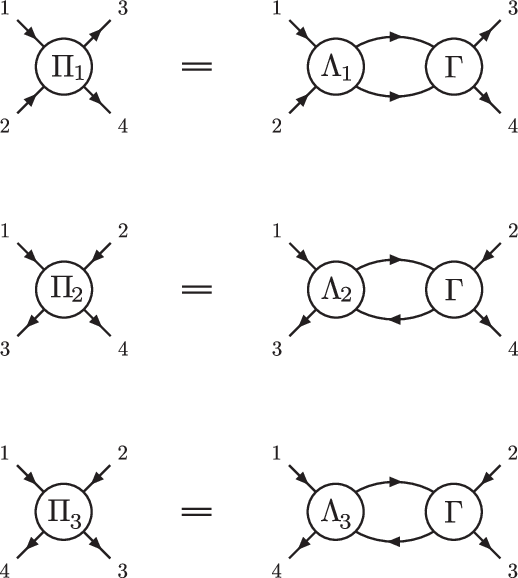}
\caption{Figure defining the parquet diagrams. The top row consisting of the figures on $\Pi_1$ defines the particle-particle channel, while the remaining two rows represent the two particle-hole channels. }
\label{fig.4}
\end{figure}

\section{Parquet Diagram Resummation}

To obtain an expression of $\Gamma({\bf k})$ as accurately as possible, the parquet resummation technique will be used hereafter following Ref.18. In this approach \cite{YM,Tsuneto}, the nonparquet diagrams illustrated in Fig.3 and any higher order ones composed of them are neglected. Just like the diagrammatic analysis for the Fermi liquid theory \cite{AGD}, the remaining diagrams, i.e., the bare vertex carrying $v_{\bf k}$ and the parquet diagrams belonging to Fig.4, will be taken into account to obtain a fully-renormalized interaction vertex function $\Gamma$ as accurately as possible. It will be shown below how the formal results derived in Ref.18 are reproduced in the present analysis using the Landau gauge (\ref{Lgauge}). 

First, the clean limit with no disorder will be considered. As sketched in Fig.4, the parquet diagrams are composed of one particle-particle channel $\Pi_1$ and the two particle-hole channels $\Pi_j$ ($j=2$ and $3$). 

The interaction term of the Hamiltonian expressed by the vertex $\Pi_j$ takes the form 
\begin{equation}
\sum_{k_1} \sum_{p_1, p_2,p_3} \exp(i(p_2-p_3)k_1/h) \, \Pi_j(k_1, p_1-p_3) \, \varphi_{p_1}^*  \varphi_{p_2}^* \varphi_{p_3}  \varphi_{p_1+p_2-p_3}. 
\end{equation}
According to this parametrization, the first row of Fig.4 expressing the relation that $\Pi_1$ satisfies is expressed as 
\begin{eqnarray}
\sum_{k_1'} \exp\biggl(i\frac{(p_2-p_3)k_1'}{h} \biggr) \Pi_1(k_1', p_1-p_3)\!&=& \! \frac{x}{N_v} \!\! \sum_{k_1', k_1^{''}, p'} \exp(i(p_2-p') k_1')/h) \, \Lambda_1(k_1', p_1-p') \nonumber \\ 
&\times& \!\! \exp\biggl(i\frac{(p_1+p_2-p_3 - p')k_1''}{h} \biggr) \, \Gamma(k_1'', p'- p_3), 
\label{Pi1}
\end{eqnarray}
where 
\begin{eqnarray}
\Gamma({\bf k}) &=& v_{\bf k} + \sum_{j=1,2,3} \Pi_j({\bf k}), \nonumber \\
\Lambda_i({\bf k}) &=& v_{\bf k} + \sum_{j \neq i} \Pi_j({\bf k}). 
\end{eqnarray} 
To change eq.(\ref{Pi1}) to a simpler form, the wave numbers $p_1-p_3$, $p_2-p_3$, and $p'-p_3$ are rewritten as $k_2$, $p$, and $k_2'$, respectively, and the resulting eq.(\ref{Pi1}) will be multiplied by the factor $\exp(-i pk_1/h)$. Further, by summing over $p$, one obtains  
\begin{equation}
\Pi_1({\bf k}) = - \frac{x}{N_v} \sum_{{\bf k}'} \Lambda_1({\bf k}-{\bf k}') \Gamma({\bf k}') \exp(i({\bf k} \times {\bf k}')_z/h) \equiv -x(\Lambda_1 
\bullet \Gamma)({\bf k}). 
\label{parquet1}
\end{equation} 

It is much easier to obtain the corresponding expressions to be satisfied by $\Pi_2$ and $\Pi_3$. Consequently, the second and third rows of Fig.4 are expressed in the form 
\begin{eqnarray}
\Pi_2({\bf k}) &=& -2x \Lambda_2({\bf k}) \Gamma({\bf k}), \nonumber \\
\Pi_3({\bf k}) &=& - \frac{2x}{N_v} \sum_{{\bf k}'} \Lambda_3({\bf k}-{\bf k}') \Gamma({\bf k}') \nonumber \\
&\equiv& - 2x (\Lambda_3 \, * \, \Gamma)({\bf k}). 
\end{eqnarray}

The formulation explained above is straightforwardly extended to the disordered case. For instance, one has only to assign replica indices $a$, $b$, $c$, and $d$ to the external propagators $1$, $2$, $3$, and $4$ in Fig.4, respectively. Then, we find 
\begin{eqnarray}
\Pi_{1, {ab,cd}}({\bf k}) &=& - x \sum_{e,f=1}^n (\Lambda_{1, {ab,ef}} \bullet \Gamma_{ef,cd})({\bf k}), \nonumber \\
\Pi_{2, {ab,cd}}({\bf k}) &=& - 2x \sum_{e,f=1}^n \Lambda_{2, {ae,cf}}({\bf k}) \Gamma_{fb,ed}({\bf k}), \nonumber \\
\Pi_{3, {ab,cd}}({\bf k}) &=& - 2x \sum_{e,f=1}^n (\Lambda_{3, {ae,fd}} \, * \, \Gamma_{fb,ce})({\bf k}), \nonumber \\
\Gamma_{ab,cd}({\bf k}) &=& \delta_{ac} \delta_{bd}(\delta_{ab} - \theta(t))v_{\bf k} + \sum_{i=1}^3 \Pi_{i, {ab,cd}}({\bf k}), \nonumber \\
\Lambda_{i, {ab,cd}}({\bf k}) &=& \Gamma_{ab,cd}({\bf k}) - \Pi_{i, {ab,cd}}({\bf k}). 
\label{parquetcomplex}
\end{eqnarray}
Further, using symmetrical relations to be satisfied between $\Pi_{i, {ab,cd}}$ and $\Lambda_{i, {ab,cd}}$ (see Appendix A), these vertex functions are represented in the form 
\begin{eqnarray}
\Pi_{1, abcd}({\bf k}) &=& \frac{1}{2} \delta_{ac} \delta_{bd} [ \delta_{ab} \Gamma_1({\bf k}) - \theta \Xi_1({\bf k}) ] \nonumber \\ 
&+& \frac{1}{2} \delta_{ad} \delta_{bc} [ \delta_{ab} {\hat \Gamma}_1({\bf k}) 
- \theta {\hat \Xi}_1({\bf k}) ], \nonumber \\
\Pi_{2, abcd}({\bf k}) &=& {\hat \Pi}_{3, abdc}({\bf k}) = \frac{1}{2} \delta_{ac} \delta_{bd} [ \delta_{ab} \Gamma_2({\bf k}) 
- \theta \Xi_2({\bf k}) ] \nonumber \\ 
&+& \frac{1}{2} \delta_{ad} \delta_{bc} [ \delta_{ab} {\hat \Gamma}_3({\bf k}) - \theta {\hat \Xi}_3({\bf k}) ], \nonumber \\
\Lambda_{1, abcd}({\bf k}) &=& \frac{1}{2} \delta_{ac} \delta_{bd} [ \delta_{ab} I_1({\bf k}) - \theta J_1({\bf k}) ] \nonumber \\ 
&+& \frac{1}{2} \delta_{ad} \delta_{bc} [ \delta_{ab} {\hat I}_1({\bf k}) 
- \theta {\hat J}_1({\bf k}) ], \nonumber \\
\Lambda_{2, abcd}({\bf k}) &=& {\hat \Lambda}_{3, abdc}({\bf k}) = \frac{1}{2} \delta_{ac} \delta_{bd} [ \delta_{ab} I_2({\bf k}) \nonumber \\
&-& \theta J_2({\bf k}) ] + \frac{1}{2} \delta_{ad} \delta_{bc} [ \delta_{ab} {\hat I}_3({\bf k}) - \theta {\hat J}_3({\bf k}) ]. 
\end{eqnarray}
Then, these representations will be substituted into eq.(37). By performing the summations on the replica indices and by taking the $n \to 0$ limit, we obtain the following forms independent of the replica indices: 
\begin{eqnarray}
\Gamma_1({\bf k}) &=& -x (I_1 \bullet f_R - \theta I_1 \bullet v_R - \theta J_1 \bullet f_R)({\bf k}), \nonumber \\
\Gamma_2({\bf k}) &=& -x (I_2({\bf k}) f_R({\bf k}) + I_2({\bf k}) {\hat h}_R({\bf k}) \nonumber \\
&+& [ {\hat I}_3({\bf k}) - \theta {\hat J}_3({\bf k})] f_R({\bf k})), \nonumber \\
\Gamma_3({\bf k}) &=& -x (I_2 \, * \, f_R - \theta I_3 \, * \, v_R - \theta J_3  \, * \, f_R)({\bf k}),
\end{eqnarray}
and
\begin{eqnarray}
\Xi_1({\bf k}) &=& x \theta(J_1 \bullet v_R)({\bf k}), \nonumber \\
\Xi_2({\bf k}) &=& -x(I_2({\bf k}) v_R({\bf k}) + J_2({\bf k})[f_R({\bf k}) + {\hat h}_R({\bf k})] \nonumber \\
&+& [{\hat I}_3({\bf k}) - \theta {\hat J}_3({\bf k})]v_R({\bf k})), \nonumber \\
\Xi_3({\bf k}) &=& x \theta(J_3 \, * \, v_R)({\bf k}),
\end{eqnarray}
where 
\begin{eqnarray}
f_R({\bf k}) &=& v_{\bf k} + \sum_{j=1}^3 \Gamma_j({\bf k}), \nonumber \\
v_R({\bf k}) &=& v_{\bf k} + \sum_{j=1}^3 \Xi_j({\bf k}), \nonumber \\
I_i({\bf k}) &=& v_{\bf k} + \sum_{j \neq i} \Gamma_j({\bf k}), \nonumber \\
J_i({\bf k}) &=& v_{\bf k} + \sum_{j \neq i} \Xi_j({\bf k}). 
\end{eqnarray}

\section{Approximate treatment of 3D case} 

The analysis in sec.2 and 3 holds only in the 2D SC films. It is quite hard to directly extend the approach in the preceding sections to the 3D case in a numerically tractable way because of the additional appearance of the component parallel to the magnetic field of the wavevector of the SC fluctuation. Nevertheless, by assuming that the nature of freezing to the vortex lattice is essentially the same as each other in both dimensions, it is not difficult to give an approximate expression of the fluctuation conductivity in 3D case. In this section, the method we have used to obtain the 3D conductivity will be explained. 

The simplest extension of the 2D GL Hamiltonian (\ref{2dgl}) is 
\begin{equation}
{\cal H}^{(3)}[\psi] = \xi_0^{-3} \!\! \int d^3r \, \biggl[ (t-1)\rho({\bf r}) + u({\bf  r})\rho({\bf  r}) + \pi t \sqrt{b^{(3)}_G} |\rho({\bf  r})|^2 + \xi_0^2 \, \biggl| \biggl( -i \nabla+\frac{2\pi}{\phi_0}{\bf A}({\bf r}) \biggr) \psi({\bf r}) \biggr|^2  \biggr], 
\end{equation}
where $\psi = \Delta \sqrt{\xi_0^3 N(0)/k_{\rm B} T}$, $\rho=|\psi|^2$, and $b_{\rm G}^{(3)} = [ g_0 k_{\rm B} T_{c0}/(2 \pi \xi_0^3 N(0)) ]^2$. In place of eq.(\ref{impcor}), the random-average is defined by $\overline{u({\bf r})} = 0$ and 
\begin{equation} 
\overline{u({\bf r}) u({\bf r}')} = \xi_0^3 \Delta^{(3)} \delta^{(3)}({\bf r}-{\bf r}'). 
\label{impcor3d}
\end{equation}
The order parameter field $\psi$ in 3D case is expanded via the product of the LLL eigenfunction $U_p$ and the plain wave propagating along the field in the manner 
\begin{equation}
\psi({\bf r}) = {\tilde L}_z^{-1/2} \sum_{p,q} \varphi_{p,q} e^{iqz} U_p(x,y), 
\end{equation}
where ${\tilde L}_z=L_z/\xi_0$. Then, the expression corresponding to eq.(\ref{LLL2dgl}) is 
\begin{eqnarray}
{\cal H}^{(3)}[\psi] &=& \sum_{p,q} (\mu_0 + q^2) |\varphi_{p,q}|^2 + \frac{t \pi \sqrt{b_{\rm G}^{(3)}}}{{\tilde S} {\tilde L}_z} \sum_{\bf K} \sum_{p,p',q,q'} \exp(i(p-p')k_1/h) \, v_{{\bf k}} \, \varphi_{p_-, q_-}^* \varphi_{p'_+, q'_+}^* \varphi_{p_+, q_+} \varphi_{p'_-, q'_-} \nonumber \\
&+& \frac{1}{\sqrt{{\tilde S} {\tilde L}_z}} \sum_{p,q,{\bf K}} u_{-{\bf K}} v_{{\bf k}}^{1/2} \exp(i p k_1/h) \varphi_{p_-, q_-}^* \varphi_{p_+, q_+}, 
\label{bareGL3d}
\end{eqnarray}
where $q_\pm=q \pm k_3/2$. and ${\bf K} = {\bf k} + k_3 \, {\hat z}$. In contrast to the 2D case, however, the coefficient of the quadratic term of the renormalized Hamiltonian corresponding to the bare one eq.(\ref{bareGL3d}) is a $q$-dependent function $\mu(q)$, and the renormalized vertex $\Gamma$ corresponding to the bare one $v_{{\bf k}}$ should also become dependent on $q$ and $q'$. It is not easy to follow such $q$-dependences entangled with the ${\bf k}$-dependences. For this reason, a general analysis for the 3D case becomes numerically formidable. 

Here, as a numerically tractable approximation, the fully renormalized vertex $\Gamma$ will be assumed hereafter not to be changed from the 2D result obtained in the manner mentioned in the preceding section. This physically means that the fluctuation effect on the vortex lattice formation is overestimated in the present approximation for 3D systems. In this approximation, eq.(\ref{parquet1}) is simply replaced by 
\begin{equation}
\Pi^{(3)}_1({\bf k}) 
= - x^{(3)} (\Lambda^{(3)}_1 \bullet \Gamma^{(3)})({\bf k}), 
\label{pi33d}
\end{equation}
where $x^{(3)} = t \, h \, \sqrt{b_G^{(3)}} \sum_q [{\cal G}(q)]^2/{\tilde L}_z$, and ${\cal G}(q)$ is the SC fluctuation propagator in LLL. Here and below, the quantity $C^{(3)}$ implies the corresponding expression in 3D case of the quantity $C$ in 2D case. Within the present approximation, other relations associated with the parquet diagrams in 3D case are also obtained in the same way, i.e., simply by replacing $x$ and $\theta$ in the corresponding ones in 2D case with $x^{(3)}$ and $\theta^{(3)}$, respectively, where 
\begin{equation}
\theta^{(3)} = \frac{\Delta^{(3)}}{2 \pi t \sqrt{b_G^{(3)}}}. 
\end{equation}

Further, the $q$-dependence in ${\cal G}(q)$ will also be assumed not to be renormalized in a consistent way with the above procedure for $\Gamma$. A similar treatment for the SC fluctuation propagator ${\cal G}$ was used previously \cite{IOT90}. Then, we have $[{\cal G}(q)]^{-1} = \mu^{(3)} + q^2$ with 
\begin{equation}
\mu^{(3)} = t-1+h+\Sigma^{(3)}. 
\label{rmass3}
\end{equation}
Then, the $q$-integrals appearing in the 3D counterpart of Fig.1 are easily performed, and the self-energy $\Sigma^{(3)}$ becomes 
\begin{equation}
\Sigma^{(3)} = 2 \mu^{(3)} x^{(3)} (2 - \theta^{(3)}) - \frac{4 \mu^{(3)} (x^{(3)})^2 (1-\theta^{(3)})}{3h} \int_0^\infty d({\bf k}^2) v_{\bf k} \Gamma^{(3)}({\bf k}), 
\end{equation}
where
\begin{equation}
x^{(3)} = \frac{t \, h \, \sqrt{b_G^{(3)}}}{4 (\mu^{(3)})^{3/2}}, 
\end{equation}  
and $\Gamma^{(3)}$ is the fully-renormalized vertex function corresponding to $\Gamma$ in 2D case and is determined from the parquet equations including eq.(\ref{pi33d}) constructed in the manner explained above. 

As already mentioned, it is anticipated that the present approximation underestimates the 3D SC ordering, i.e., the growth of the lateral correlation on the vortex position, compared with that to be obtained from a more realistic method for 3D case. Nevertheless, it will be seen in sec.4 that the resistivity in 3D case obtained numerically is strongly affected by the growth of the positional correlation length and consequently that our view on the vanishing of the resistivity in 3D case becomes clear. 

\section{Vortex Glass Correlation Function} 

In order to derive the VG correlation function $G_{\rm vg}$ and the resulting term of the conductivity accompanied by $G_{\rm vg}$, the starting GL model in sec.2 will be rewritten as a quantum action by giving a purely dissipative dynamics to the order parameter. In the unit $\hbar=1$, the partition function $Z$ in $D$-dimensional case is expressed as $Z= {\rm Tr} \, \exp(-{\cal S}_{\rm GL})$ using the quantum GL action \cite{Brezin,Abrahams,IOT91} 
\begin{equation}
{\cal S}_{\rm GL} = \xi_0^{-D} \int d^D{\bf r} \sum_\omega \, \gamma |\omega| \, \psi^*({\bf r}, \omega) \psi({\bf r}, \omega) + T \int_0^{1/T} d\tau \, {\cal H}[\psi(\tau)], 
\label{2dGLaction}
\end{equation}
where ${\cal H}[\psi]$ in $D=2$ case was given as eq.(\ref{r2dgl}), $\omega$ is the Matsubara frequency for bosons, $\tau$ is the imaginary time, $\gamma > 0$, and 
the Fourier transform $\psi_\omega$ of $\psi(\tau)$ is defined by 
\begin{equation}
\psi(\tau) = \sum_\omega \psi_\omega e^{-i \omega \tau}. 
\end{equation}
It is expected, close to the VG transition, that the dominant frequency dependence determining the time scale of the conductivity is carried by the VG critical fluctuation and that detailed frequency dependences of the self-energy of the noncritical SC fluctuation will play less important roles. For this reason, the frequency dependence of the fully renormalized vertex part $\Gamma$ representing the positional correlation between the vortices is assumed to be negligible. This assumption should be valid in clean enough cases of our interest. Next, the frequency dependence of the SC fluctuation propagator ${\cal G}$ will be kept the bare one. That is, the bare form ${\cal G}=1/(\mu+\gamma|\omega|)$ will be always used in 2D case. 

\begin{figure}
\includegraphics{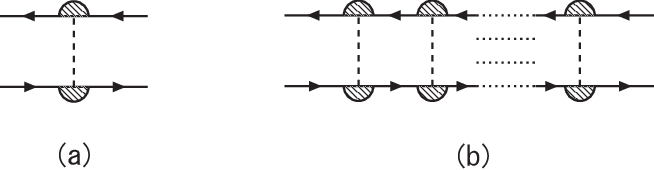}
\caption{Structure of the vortex-glass correlation function $G_{\rm vg}$ which is represented in the ladder approximation valid at weak disorder. Figure (a) is the four-point vertex constructed based on the hatched semicircles of Fig.2 after the random average. Here, the connected dashed line denotes a pair-wised random potential occurring after the random-average and carrying $\Delta^{(2)}$ in 2D case. As shown in (b), the VG correlation function is expressed as a series of the figure (a). }
\label{fig.5}
\end{figure}

The VG correlation function is given by 
\begin{equation}
G_{\rm vg}({\bf R}; \tau_1, \tau_2) \propto \int d^D{\bf r} \, \overline{|\langle \psi^*({\bf r}, \tau_1) \psi({\bf r}+{\bf R}, \tau_2) \rangle|^2}. 
\label{vgcorr}
\end{equation}

For a while, we focus on the 2D ($D=2$) case. The genuine VG transition, i.e., the 2D SC transition, does not occur in real systems at finite temperatures in a field perpendicular to the plane \cite{FFH,DHF}. Nevertheless, the VG fluctuation in 2D case will be considered here in the Gaussian approximation where the VG transition line inevitably appears, in order to make a comparison between the 2D and 3D cases useful. 
The Fourier transform of eq.(\ref{vgcorr}) with $D=2$ becomes 
\begin{equation}
G_{\rm vg}({\bf k}; \omega_1, \omega_2) = \frac{1}{N_v} [{\cal G}(\omega_1) {\cal G}(\omega_2)]^{-1} \sum_{p,p'} \exp(i(p-p')k_1/h) \, {\overline{{\cal G}_{p',p}(\omega_1) {\cal G}_{p+k_2, p'+k_2}(\omega_2)}} 
\label{vgcorr2d}
\end{equation}
in a form normalized properly, where ${\cal G}_{p,p'}(\omega) = \langle \varphi_p(\omega) \varphi^*_{p'}(\omega) \rangle$ is the LLL fluctuation propagator defined prior to taking the random average, and ${\cal G}(\omega)= \overline{{\cal G}_{p,p'}(\omega)} = 1/(\mu + \gamma |\omega|)$ is the random-averaged LLL fluctuation propagator. First, the irreducible disorder vertex forming ${\cal G}_{p,p'}(\omega)$ will be explained. As far as the limit of weak disorder is concerned, it is sufficient to regard the diagram with a hatched semicircle of Fig.2 as the irreducible one for ${\cal G}_{p,p'}(\omega)$. By incorporating the frequency dependences in the SC fluctuation propagators of Fig.2, the corresponding expression to eq.(\ref{lambdazero}) becomes 
\begin{eqnarray}
\Lambda(p,p'; \omega_1) &=& {\tilde S}^{-1/2} [{\cal G}(\omega_1)]^2 \sum_{\bf k} \delta_{p, p'+k_2} \, u_{-{\bf k}} \, v_{\bf k}^{1/2} \exp(ik_1(p+p')/2h) \nonumber \\ 
&\times& \biggl[ 1 - \frac{2 x \mu^2}{N_v} \sum_{{\bf k}', {\omega_1}'} [{\cal G}(\omega'_1)]^2 \, \Gamma({\bf k}') \exp(i(k_1'(p'-p) - k_2' k_1)/h) \biggr].
\end{eqnarray}
\label{ppp}
Since the dependences on the internal frequency $\omega'_1$ are unnecessary in obtaining the conductivity in the regime where the quantum SC fluctuation is negligible, however, only the $\omega'_1=0$ term will be kept hereafter. Then, the expression in the square bracket of eq.(55) simply becomes 
\begin{equation}
f(p-p', k_1) = 1 - 2 \, x \, \Gamma(k_1, p-p'), 
\label{vconaji}
\end{equation}
and the diagram of Fig.5 (a) arising after random -averaging takes the form 
\begin{equation}
\frac{\Delta^{(2)}}{\tilde S} [{\cal G}(\omega_1) {\cal G}(\omega_2)]^2 \sum_{\tilde {\bf p}} \delta_{{\tilde p}_2, p-p'} \, v_{\bf {\tilde p}} \, \exp(-ik_2 {\tilde p}_1/h) \, 
 [f(p-p', {\tilde p}_1)]^2, 
\label{irrvg}
\end{equation}
where ${\bf {\tilde p}}=({\tilde p}_1, {\tilde p}_2)$. Note that eq.(\ref{vconaji}) is the same expression as eq.(\ref{vcorigin}). Taking account of the fact that the VG correlation function $G_{\rm vg}$ is expressed as a ladder series Fig.5 (b) by regarding the expression (\ref{irrvg}) as its unit, it is found that 
\begin{equation}
G_{\rm vg}({\bf k}; \omega, \omega+\Omega) = \frac{1}{1 - X({\bf k}) {\cal G}(\omega) {\cal G}(\omega+\Omega)}, 
\end{equation}
where 
\begin{equation}
X({\bf k}) \equiv \frac{\mu^2 x \theta}{N_v} \sum_{\bf p} v_{\bf p} |f({\bf p})|^2 \, \exp(i {\bf k}\cdot{\bf p}/h). 
\end{equation}
Based on this expression of $G_{\rm vg}$, the VG correlation length $\xi_{\rm vg}$ just above the VG transition is expressed by 
\begin{equation}
\xi_{\rm vg} = \frac{1}{2} \sqrt{\frac{c_2}{(1 - c_0)h}}, 
\label{vglength}
\end{equation}
where 
\begin{eqnarray}
c_0 &=& \frac{x \theta}{N_v} \sum_{\bf p} v_{\bf p} |f({\bf p})|^2, \nonumber \\
c_2 &=& \frac{x \theta}{h N_v} \sum_{\bf p} {\bf p}^2 v_{\bf p} |f({\bf p})|^2, 
\end{eqnarray}
and $\xi_0$ is the unit of the length scales. Of course, eq.(\ref{vglength}) is valid when $c_0 < 1$ and diverges on approaching the VG transition point at which $c_0 = 1$.

\subsection{3D case}

As in 2D case, the 3D VG transition will be described below in the Gaussian fluctuation. 
First, let us start from determining the Fourier transform of eq.(\ref{vgcorr}) properly. By choosing the prefactor of eq.(\ref{vgcorr}) in a consistent manner with eq.(\ref{vgcorr2d}), the expression in 3D corresponding to eq.(\ref{vgcorr2d}) is given by 
\begin{eqnarray}
G_{\rm vg}({\bf K}; \omega_1, \omega_2) &=& \frac{1}{N_v} \biggl( {\sum_q {\cal G}(q,\omega_1) {\cal G}(q+k_3, \omega_2)} \biggr)^{-1} \sum_{p,p'; q,q'} [ \, \exp(i(p-p')k_1/h) \nonumber \\
&\times& \overline{\langle \varphi^*_{p,q}(\omega_1) \varphi_{p',q'}(\omega_1) \rangle \langle \varphi_{p+k_2,q+k_3}(\omega_2) \varphi^*_{p'+k_2,q'+k_3}(\omega_2) \rangle} \, ], 
\end{eqnarray}
where ${\bf K}= {\bf k} + k_3 {\hat z}$, ${\bf k}=(k_1, k_2)$, and ${\cal G}(q,\omega)$ is the random averaged fluctuation propagator in 3D case. The expression in 3D case corresponding to eq.(51) is 
\begin{eqnarray}
\Lambda(p,p'; q,q'; \omega_1)\!&=& \! ({\tilde S} {\tilde L}_z)^{-1/2} [{\cal G}(q,\omega_1) {\cal G}(q', \omega_1)] \sum_{\bf {\tilde P}} \delta_{p,p'+{\tilde p}_2} \delta_{q,q'+{\tilde p}_3} \,  u_{-{\bf {\tilde P}}} \, v_{\bf {\tilde p}}^{1/2} \exp(i(p+p')p_1/2h) \nonumber \\
&\times& \biggl[ 1 - \frac{8 (\mu^{(3)})^{3/2} x^{(3)}}{{\tilde L}_z N_v} \sum_{{\bf K}', \omega'_1} [{\cal G}(q+k'_3, \omega'_1) {\cal G}(q'+k'_3, \omega'_1)]
\nonumber \\
&\times&  \Gamma^{(3)}({\bf k}') \exp(i[(p'-p)k'_1 
- {\tilde p}_1 k'_2]/h) \biggr], 
\end{eqnarray}
where ${\bf {\tilde P}}= {\bf {\tilde p}} + p_3 {\hat z}$, and ${\bf {\tilde p}}=({\tilde p}_1, {\tilde p}_2)$. 
By neglecting the dependences on the internal frequency $\omega'_1$ in $\Lambda(p, q; p',q'; \omega_1)$, the vertex correction corresponding to $f(p-p', {\tilde p}_1)$ in 2D case becomes 
\begin{equation}
f^{(3)}(p-p', {\tilde p}_1; q-q') \equiv 1 -  \frac{8 x^{(3)} \mu^{(3)}}{4 \mu^{(3)} + (q-q')^2} \Gamma^{(3)}(p-p', {\tilde p}_1), 
\label{impvc3d}
\end{equation}
and the expression representing the counterpart of Fig.5 (a) is 
\begin{equation}
\frac{\Delta^{(3)}}{{\tilde S} {\tilde L}_z} \sum_{{\bf K}'} \delta_{k'_2, p-p'} \delta_{k'_3, q-q'} v_{{\bf k}'} [f^{(3)}(k'_1, p-p'; q-q')]^2 e^{-i \frac{k_2 k'_1}{h}} {\cal G}(q,\omega_1) {\cal G}(q', \omega_1) {\cal G}(q+k_3, \omega_2) {\cal G}(q'+k_3, \omega_2). 
\end{equation} 
Here, consistently with the approximation for $\Sigma^{(3)}$, the $q-q'$-dependence in eq.(\ref{impvc3d}) will be neglected. Then, it is straightforward to derive the expression on the Fourier transform of the VG correlation function in 3D, and we have 
\begin{equation}
G^{(3)}_{\rm vg}({\bf K}; \omega, \omega + \Omega) = \biggl[ 1 - \frac{X^{(3)}({\bf k})}{{\tilde L}_z} \sum_q {\cal G}(q,\omega) {\cal G}(q+k_3, 
\omega+\Omega) \biggr]^{-1}, 
\label{vgcorr3d}  
\end{equation}
where
\begin{equation}
X^{(3)}({\bf k}) = \frac{4 (\mu^{(3)})^{3/2} x^{(3)} \theta^{(3)}}{N_v} \sum_{{\bf k}'} v_{{\bf k}'} [f^{(3)}({\bf k}'; 0)]^2 \exp(i({\bf k} \times {{\bf k}'})_z/h). 
\end{equation}

As in 2D case, the vanishing behavior of the resistivity, which will be discussed in the next section, is expected to be determined by the time scale not of the SC fluctuation but of the VG correlation function. For this reason, the frequency term of ${\cal G}^{-1}_\omega$ as well as the wave number dependence in 3D case will be assumed to be unrenormalized by the randomness. Thus, the expression \begin{equation}
{\cal G}(q,\omega) = \frac{1}{\mu^{(3)} + q^2 + \gamma|\omega|}
\label{appdynG}
\end{equation}
for the SC fluctuation propagator with the wave number $q$ will be used hereafter in 3D case. However, it will be discussed later that, in dirty cases, this assumption has an unfavorable effect on the vanishing of the resistivity. 

Then, by focusing on the vanishing behavior of the r.h.s. of eq.(\ref{vgcorr3d}), the VG correlation lengths are defined in a similar manner to that in 2D case. The VG correlation length $\xi_{{\rm vg}, \perp}$ in perpendicular directions to the magnetic field is given by 
\begin{equation}
\xi_{{\rm vg}, \perp} = \frac{1}{2} \sqrt{\frac{c_2^{(3)}}{(1 - c_0^{(3)})h}}, 
\label{xiperp}
\end{equation}
while the corresponding correlation length in the direction of the magnetic field is 
\begin{equation}
\xi_{{\rm vg},\parallel} = \sqrt{\frac{h}{c_2^{(3)} \mu^{(3)}}} \, \, \xi_{{\rm vg}, \perp}, 
\label{xipara}
\end{equation}
where 
\begin{eqnarray}
c_0^{(3)} &=& \frac{x^{(3)} \theta^{(3)}}{N_v} \sum_{\bf p} v_{\bf p} |f^{(3)}({\bf p}; 0)|^2, \nonumber \\ 
c_2^{(3)} &=& \frac{x^{(3)} \theta^{(3)}}{h N_v} \sum_{\bf p} {\bf p}^2 v_{\bf p} |f^{(3)}({\bf p}; 0)|^2. 
\end{eqnarray}

\section{Vortex Glass Contribution to Conductivity} 

In general, the vortex pinning effect is formulated according to the two types of modellings. One is the so-called random $T_c$ model corresponding to the inclusion of the scalar potential $u({\bf r})$ in eq.(\ref{2dgl}), and what is pinned in this case is a spatial variation of the amplitude $|\psi| = \sqrt{\rho}$ associated with the vortex core. However, this model is not useful in an approach based on the London, or a phase-only model where $\rho$ is a constant. The other model useful even in the phase-only approach is the following one expressed in terms of a random gauge potential model: In 2D case, the term 
\begin{equation}
\delta {\cal S}_g = T \int_0^{1/T} d\tau \int d^2{\bf r} \,\,\, {\bf j}({\bf r}, \tau)\cdot{\bf a}_r({\bf r}), 
\end{equation}
should be added to eq.(\ref{2dGLaction}), where ${\bf j} = \psi^* (-i\nabla + 2 \pi {\bf A}/\phi_0 ) \psi + {\rm {c.c.}}$ is the supercurrent density. Using the fact that, due to the continuity equation $\nabla\cdot{\bf j}=0$ to be satisfied in equilibrium, ${\bf a}_r$ may be rewritten as $\nabla \times w({\bf r}) {\hat z}$ by introducing a scalar function $w({\bf r})$, the total random potential term ${\cal S}_r$ may be reexpressed as 
\begin{equation}
\delta {\cal S}_r = \int_0^{1/T} d\tau \int d^2{\bf r} \, [ w({\bf r}) [\nabla \times {\bf j}({\bf r}, \tau)]_z + \xi_0^{-2} u({\bf r}) \, \rho({\bf r}, 
\tau)]. 
\label{randomaction+}
\end{equation}
The random potential $w$ is assumed to yield 
${\overline {w({\bf r})}}$=0, and 
\begin{equation}
\overline{w({\bf r}) w({\bf r}')} = \Delta_\Phi \xi_0^2 \delta^{(2)}({\bf r} - {\bf r}')  
\end{equation}
in 2D case. Since, under a fixed $\rho$, $w({\bf r})(\nabla \times {\bf j})_z$ is $- 2 \pi \rho w({\bf r}) \xi_0^2 (n_v({\bf r}) - H/\phi_0)$, $w({\bf r})$ can be regarded as the pinning potential for vortex cores \cite{Natter,GleD}, where $n_v({\bf r})$ is the vortex density. As in the previous works \cite{RI96,RI962,RIMyojin} based on the GL approach, the randomness on $T_c$, i.e., $\Delta^{(2)}$ or $\Delta^{(3)}$, is assumed to be much larger than $\Delta_\Phi$. 

To derive the conductivity terms associated with the current due to the SC fluctuations, we follow the method \cite{IOT89,DHF,RI95,RI962,RIMyojin,II02} based on the use of the Kubo formula. 
In the case under an applied current perpendicular to the magnetic field, the second (or, $n=1$) LL modes are needed \cite{IOT89} in addition to the LLL modes to obtain the fluctuation conductivity under a uniform current. By keeping only the 0th and 1st order terms with respect to the SC fluctuation $\varphi_1$ belonging to the $n=1$ LL, the action (\ref{randomaction+}) is written as 
\begin{eqnarray}
{\cal S}_r &=& {\tilde S}^{-1} \sum_{\omega, p, {\bf k}} v_{\bf k}^{1/2} \exp(i pk_1/h) \biggl[ (u_{-{\bf k}} - {\bf k}^2 w_{-{\bf k}}) \, \varphi^*_{0,p_-}(\omega) \varphi_{0, p_+}(\omega) \nonumber \\
&-& \frac{i}{\sqrt{2h}} [u_{-{\bf k}} +(2h - {\bf k}^2) w_{-{\bf k}}](k_+ \varphi^*_{0,p_-}(\omega) \varphi_{1, p_+}(\omega) + k_- \varphi^*_{1,p_-}(\omega) \varphi_{0, p_+}(\omega) ) \biggr], 
\end{eqnarray}
where $p_\pm = p \pm k_2/2$, and $k_\pm = k_1 \pm i k_2$. For simplicity, we focus on a high field range in which the $n=1$ LL mode is so heavy and hence, assume that the mass $\mu_1$ of the $n=1$ LL fluctuation remains its renormalized value \cite{RI95} $2h$ and that its dynamics is negligible. 

The Kubo formula of the SC part of the dc conductivity under a current in the $x$-direction is expressed in the form 
\begin{equation}
\sigma_{s, xx} = - k_{\rm B} T \lim_{\Omega \to +0} \frac{\partial}{\partial \Omega} \, \frac{\delta^2 \overline{{\rm log} \langle \exp(-{\cal S}_{tot}({\bf A}')) \rangle}}{\delta A'_x(0, i\Omega) \delta A'_x(0, -i\Omega)} \biggl|_{{\bf A}'=0},
\end{equation}
where ${\cal S}_{tot} = {\cal S}_{\rm GL} + \delta S_g$, and the gauge substitution ${\bf A} \to {\bf A} + {\bf A}'$ was performed to induce the current vertices. By keeping only the terms consisting of the $n=0$ and $1$ LLs, the AL and VG terms of the SC part $\sigma_{s, xx}$ of the conductivity in 2D case arise from the expression 
\begin{equation}
\frac{d R_q \sigma_{s,xx}}{4 h^2} = - \frac{k_{\rm B} T}{N_v} \lim_{T \to \infty} \lim_{\Omega \to +0} \frac{\partial}{\partial \Omega} \sum_{\omega, \omega'}  \sum_{p,p'} {\rm Re}\overline{\langle \varphi_{0,p}(\omega) \varphi^*_{0, p'}(\omega') \varphi^*_{1, p}(\omega + \Omega) \varphi_{1,p'}(\omega' + \Omega) \rangle}, 
\end{equation}
where $R_q=\pi \hbar/(2 e^2) = 6.45$ ($k\Omega$) is the quantum of resistance, and, to simplify our analysis on $\sigma_{s,xx}$, the formal $T \to \infty$ limit is taken to neglect the {\it quantum} SC fluctuations. In the present high field approximation, the VG fluctuation is formed by the LLL SC fluctuation, while the spatially averaged current vertex is inevitably accompanied by the $n=1$ LL fluctuation. The VG fluctuation term $\sigma_{{\rm vg}, xx}$ of the conductivity is accompanied by vertices due to random potentials unrelated to the VG correlation function (see Fig.6). Below, such a random potential vertex will be called an outer random potential. 

\begin{figure}
\includegraphics{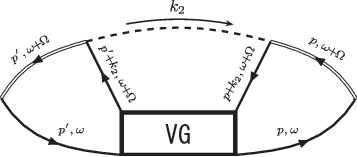}
\caption{Diagram expressing $\sigma_{1, {\rm vg},xx}^{(2)}$. The two double lines with the frequency $\omega+\Omega$ are the fluctuation propagators in the second LL. This conductivity term is accompanied by a single outer random potential line forming an interaction between the LLL and the next lowest LL 
fluctuations.}
\label{fig.6}
\end{figure}

The lowest (i.e., the first) order term in the outer random potential in 2D case, $\sigma_{1, {\rm vg}, xx}^{(2)}$, is described by Fig.6 \cite{RI96}. Its concrete expression is 
\begin{equation}
d R_q \sigma_{1, {\rm vg}, xx}^{(2)} = \frac{\Delta^{(2)} k_{\rm B} T}{4 \pi N_v} \sum_{\bf k} {\bf k}^2 v_{\bf k} \, J_1^{(2)}({\bf k}), 
\label{sigvg1}
\end{equation}
where 
\begin{eqnarray}
J_1^{(2)}({\bf k}) &=& \lim_{T \to \infty} \biggl(-\frac{\partial}{\partial \Omega} \biggr) \sum_\omega {\cal G}(\omega) {\cal G}(\omega+\Omega) \nonumber \\
&\times& G_{\rm vg}({\bf k}; \omega, \omega+\Omega) \biggr|_{\Omega \to +0}. 
\label{sigvg1j}
\end{eqnarray}
In obtaining eq.(\ref{sigvg1}), the inequality $\Delta_\Phi \ll \Delta^{(2)}$ was assumed because the main vortex pinning effect in the GL approach should be due to the variation of the amplitude $|\psi|$ of the SC order parameter. Using the relation 
\begin{equation}
J_1^{(2)}({\bf k}) \simeq [G_{\rm vg}({\bf k};0,0)]^2 \lim_{T \to \infty} \biggl(-\frac{\partial}{\partial \Omega} \biggr) \sum_\omega {\cal G}(\omega) {\cal G}(\omega+\Omega) \biggr|_{\Omega \to +0} 
\end{equation}
of which the validity is easily checked, it is found that $J_1^{(2)}({\bf k}) \simeq \gamma [G_{\rm vg}({\bf k};0,0)]^2/(2 \mu^3)$ (see Appendix B). Thus, we obtain 
\begin{equation}
d R_q \sigma_{1, {\rm vg},xx}^{(2)} \simeq \frac{\Delta^{(2)} \gamma k_{\rm B} T}{4 h \mu^3} \int_{\bf k} {\bf k}^2 v_{\bf k} [G_{\rm vg}({\bf k};0,0)]^2.
\end{equation} 
Or, by using $\xi_{\rm vg}$ defined previously, $\sigma_{1, {\rm vg},xx}^{(2)}$ finally becomes 
\begin{equation}
d R_q \sigma_{1, {\rm vg},xx}^{(2)} = \frac{\Delta^{(2)} h \gamma k_{\rm B} T}{16 \pi \mu^2 c_2^2} \biggl[ \, {\rm log}\biggl(\frac{1 + \xi_{\rm vg}^2}{e} \biggr) + \frac{1}{1+\xi_{\rm vg}^2} \biggr], 
\label{vgc21}
\end{equation}
where $\sqrt{h}$ was chosen as the cutoff of the wave number $|{\bf k}|$. 
\begin{figure}
\includegraphics{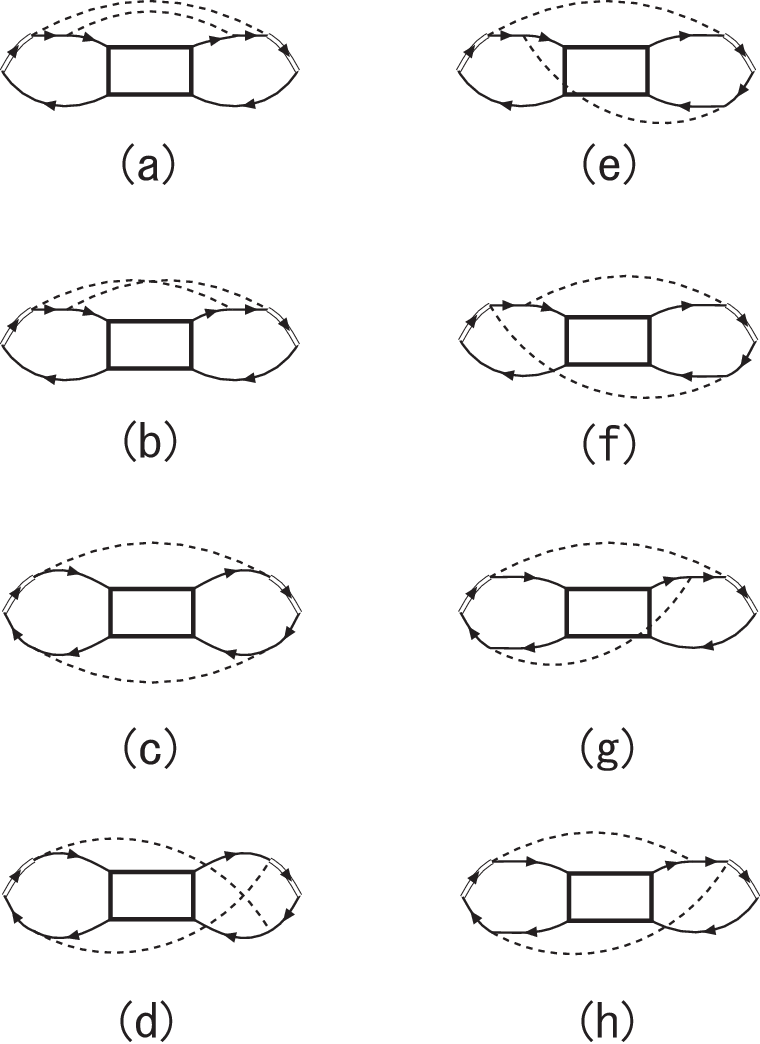}
\caption{Eight diagrams expressing $\sigma_{2,{\rm vg},xx}^{(D)}$ with $D=2$ and $3$. Here, the dashed curves are linear combinations of $\Delta^{(D)}$ and $\Delta_\Phi$ .}
\label{fig.7}
\end{figure}

The next lowest (i.e., the second) order term in the outer random potential, $\sigma_{2,{\rm vg},xx}^{(2)}$, is more divergent close to the VG transition 
point than $\sigma_{1,{\rm vg},xx}^{(2)}$ \cite{RI962,RIMyojin}. This second order term is composed of the contributions of the eight diagrams in Fig.7. As an example, let us examine Fig.7 (a) and (b) which are expressed as 
\begin{eqnarray}
d R_q \sigma_{2,{\rm vg},xx}^{(2)}|_{a+b} &=&  \frac{k_{\rm B} T}{{\tilde S} N_v} \sum_{{\bf k}, {\bf k}'} v_{\bf k} v_{{\bf k}'} \lim_{T \to \infty} \biggl(-\frac{\partial}{\partial \Omega} \biggr) \sum_\omega [{\cal G}(\omega + \Omega)]^2 {\cal G}(\omega) {\cal G}(\omega+\Omega) G_{\rm vg}({\bf k}-{\bf k}'; \omega, \omega+\Omega) \nonumber \\ 
&\times& \biggl[[\Delta^{(2)} + \Delta_\Phi(2h - {\bf k}^2)^2][\Delta^{(2)} + \Delta_\Phi {{\bf k}'}^4] {\bf k}^2 
- [\Delta^{(2)} + \Delta_\Phi {\bf k}^2 ({\bf k}^2-2h)] \nonumber \\ 
&\times& [\Delta^{(2)} + \Delta_\Phi {{\bf k}'}^2 ({{\bf k}'}^2 -2h)] k_- k'_+ \exp(i({\bf k} \times {\bf k}')_z/h) \biggr] \biggl|_{\Omega \to +0}. 
\label{ab}
\end{eqnarray}
Contributions of the remaining six diagrams can be expressed in a similar manner \cite{RIMyojin}. Close to the VG transition, however, it is sufficient to keep contributions of the lowest order in ${\bf k} - {\bf k}'$ in the coefficient corresponding to the second and third lines of eq.(\ref{ab}). By summing these contributions of the eight diagrams in this way, the full expression of $\sigma_{2,{\rm vg},xx}
^{(2)}$ becomes 
\begin{equation}
d R_q \sigma_{2, {\rm vg},xx}^{(2)} = 4 \Delta^{(2)} \Delta_\Phi h k_{\rm B} T \int_{\bf k} \int_{{\bf k}'} \frac{{\bf k}^2}{2} (v_{\bf k})^2 J_2^{(2)}({\bf k} - {\bf k}') \simeq \frac{\Delta^{(2)} \Delta_\Phi}{2 \pi} h^3 k_{\rm B} T \int_{\bf k} J_2^{(2)}({\bf k}), 
\label{sigvg2tot}
\end{equation}
where 
\begin{equation}
J_2^{(2)}({\bf q}) = \lim_{T \to \infty} \biggl(-\frac{\partial}{\partial \Omega} \biggr) \sum_\omega [{\cal G}(\omega) + {\cal G}(\omega + \Omega)]^2 {\cal G}(\omega) {\cal G}(\omega+\Omega) G_{\rm vg}({\bf q}; \omega, \omega+\Omega) \biggl|_{\Omega \to +0}. 
\end{equation}

Just like the procedure for $J_1^{(2)}$, we have evaluated $J_2^{(2)}$ in terms of the series expansion of the VG correlation function (see Appendix B). 
As the leading term of $J_2^{(2)}({\bf q})$, we obtain 
\begin{equation}
J_2^{(2)}({\bf q}) \simeq \frac{43 \gamma}{8 \mu^5} [G_{\rm vg}({\bf q}; 0,0)]^2. 
\end{equation}
Using this and the result on $\xi_{\rm vg}$, $\sigma_{2,{\rm vg},xx}^{(2)}$ in 2D case becomes 
\begin{equation}
d R_q \sigma_{2,{\rm vg},xx}^{(2)} = \frac{43 \Delta^{(2)} \Delta_\Phi \gamma k_{\rm B} T h^5}{4 \pi^2 c_2^2 \mu^5} \xi_{\rm vg}^2. 
\label{vgc22}
\end{equation}

\subsection{3D case}

As already mentioned, the fully renormalized vertex part in 3D case is assumed to have the same structure as that in 2D case. Under this simplification, it is straightforward to extend the derivation on the conductivity in 2D case explained above to that in 3D 
case. 

First, the corresponding lowest order term to eq.(\ref{sigvg1}) simply becomes 
\begin{equation}
\xi_0 R_q \sigma^{(3)}_{1,{\rm vg}, xx} = \frac{\Delta^{(3)} k_{\rm B} T}{4 \pi N_v} \int_{\bf k} {\bf k}^2 \, v_{\bf k} \int_{k_3} J^{(3)}({\bf K}), 
\label{3dsigvg1}
\end{equation}
where ${\bf K} = {\bf k} + k_3 {\hat z}$, and 
\begin{equation}
J^{(3)}({\bf K}) = \lim_{T \to \infty} \biggl(-\frac{\partial}{\partial \Omega} \biggr) \sum_\omega \int_q {\cal G}(q, \omega) {\cal G}(q+k_3, \omega+\Omega) G^{(3)}_{\rm vg}({\bf K}; \omega, \omega+\Omega) 
\biggr|_{\Omega \to +0}. 
\label{3dsigvg1j}
\end{equation}
As in 2D case, $J_3$ can be rewritten in the form proportional to $[G^{(3)}_{\rm vg}({\bf K}; 0, 0)]^2$ : 
\begin{eqnarray}
J^{(3)}({\bf K}) \simeq \frac{3 \gamma}{32 (\mu^{(3)})^{5/2}} \, [G^{(3)}_{\rm vg}({\bf K}; 0, 0)]^2 \, \frac{4 \mu^{(3)}}{4 \mu^{(3)} + k_3^2}. 
\label{J3}
\end{eqnarray}
In 3D case, however, the resulting $\sigma_{1,{\rm vg}, xx}^{(3)}$ remains nondivergent on approaching the VG transition from above and thus, will be neglected hereafter. 

The 3D conductivity term of the next order in the outer random potential can also be easily found because the wavevector dependences accompanying the outer random potential lines are decoupled to the dependences on the wavevector component parallel to the field in the fluctuation propagators. The corresponding expression in 3D case to eq.(\ref{sigvg2tot}) simply becomes 
\begin{equation}
\xi_0 R_q \sigma_{2,{\rm vg},xx}^{(3)} \simeq \frac{\Delta^{(3)} \Delta_\Phi}{2 \pi} h^3 k_{\rm B} T \int_{\bf k} J_2^{(3)}({\bf k}), 
\end{equation}
where 
\begin{equation}
J_2^{(3)}({\bf k}) = \lim_{T \to \infty} \biggl(-\frac{\partial}{\partial \Omega} \biggr) \sum_\omega {\cal A}({\bf k}; \omega, \Omega),
\end{equation}
with 
\begin{eqnarray}
{\cal A}({\bf k}; \omega, \Omega)\!&=& \! \int_{k_3} \int_{{k_3}'} \int_q {\cal G}(q, \omega) {\cal G}(q+k_3-k'_3, \omega+\Omega) {\cal G}(q-k'_3, \omega) \, \biggl[ \, {\cal G}(q-k'_3, \omega) \nonumber \\
&+& \!\! G^{(3)}_{\rm vg}({\bf k}+(k_3-k'_3){\hat z}; \omega, \omega+\Omega) \, X^{(3)}({\bf k}) \int_{q'} {\cal G}(q', \omega) \nonumber \\
&\times& {\cal G}(q'+k_3-k'_3, \omega+\Omega) {\cal G}(q'-k'_3, \omega) \, \biggr]. 
\end{eqnarray}
Here, by noting the fact that the wave numbers $q$ and $q'$ appearing in the ladder of the VG correlation function are small as well as $k_3-k'_3$, we may replace ${\cal G}(q-k'_3,\omega)$ and ${\cal G}(q'-k'_3,\omega)$ by ${\cal G}(-k'_3, 0)$. Then, we have 
\begin{equation}
J^{(3)}_2({\bf K}) \simeq \int_{k'_3} [{\cal G}(k'_3, 0)]^2 \int_{k_3} J^{(3)}({\bf K}) = \frac{1}{4 \mu^{3/2}} \int_{k_3} J^{(3)}({\bf K}). 
\end{equation}
By substituting eq.(\ref{J3}) into the above expression, $\sigma^{(3)}_{2,{\rm vg},xx}$ becomes 
\begin{equation}
\xi_0 R_q \sigma_{2,{\rm vg},xx}^{(3)} = \frac{3 \Delta^{(3)} \Delta_\Phi \gamma k_{\rm B} T h^5}{128 \pi^2 (c_2^{(3)})^2 (\mu^{(3)})^4} \, \frac{\xi_{{\rm vg}, \perp}^2}{\xi_{{\rm vg}, \parallel}}. 
\label{vgc3}
\end{equation}

\section{Numerical results of resistivity curves}

\begin{figure}
\includegraphics{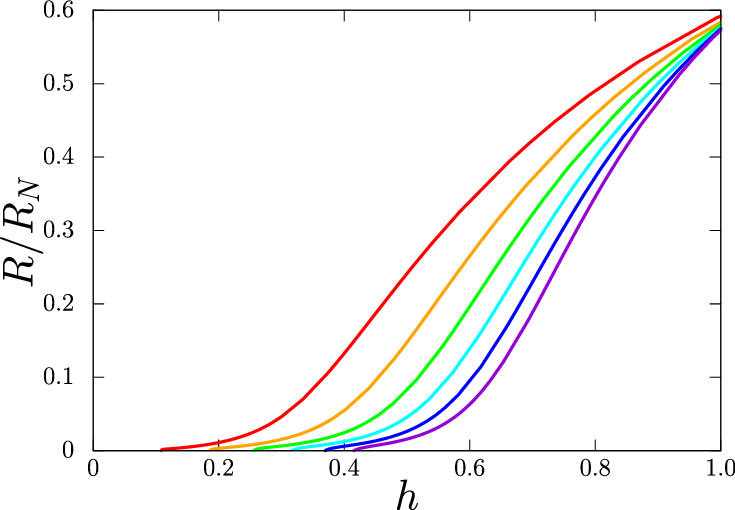}
\caption{$R/R_N$ vs $h=H/H_{c2}(0)$ curves in 2D case at $t=T/T_{c0}=0.637$ (red), $0.531$, $0.455$, $0.398$, $0.354$, and $0.318$ (violet) from left to right. The used values of the material parameters $b_G^{(2)}$ and $\Delta^{(2)}=10 \Delta_\Phi$ are $0.01$ and $0.002$, respectively. }
\label{fig.8}
\end{figure}

Using the formula on the vortex-glass (VG) contributions $\sigma_{{\rm vg},xx}$ to the conductivity obtained in sec.6, eq.(\ref{vgc3}) in 3D and eqs.(\ref{vgc21}) and (\ref{vgc22}) in 2D cases, the resulting resistivity curves will be examined in this section. The total conductivity $\sigma_{\rm tot}$ in each dimension consists of the normal conductivity $\sigma_N$, the so-called Aslamasov-Larkin (AL) term $\sigma_{{\rm AL},xx}^{(D)}$ which is essentially the SC part of the conductivity in the disorder-free case, and the VG terms. In 2D case, we set $\sigma_{\rm tot} = \sigma_{N} + \sigma_{{\rm AL},xx}^{(2)} + \sigma_{1 {\rm vg},xx}^{(2)} + \sigma_{2 {\rm vg},xx}^{(2)}$, while we set $\sigma_{\rm tot} = \sigma_{N} + \sigma_{{\rm AL},xx}^{(3)} + \sigma_{2 {\rm vg},xx}^{(3)}$ in 3D case. 

Regarding the normal term $\sigma_{N}$ of the conductivity, we simply assume that $d R_q \sigma_N = 1$ in 2D case, and $\xi_0 R_q \sigma_N = 1$ in 3D case, respectively. 

In 2D case, the AL term $\sigma_{{\rm AL},xx}^{(2)}$ is given by 
\begin{eqnarray}
d R_q \sigma^{(2)}_{{\rm AL}, xx} &=& - 4 T h^2 \biggl(\frac{\partial}{\partial \Omega} \biggr) \sum_\omega {\cal G}(\omega) \frac{1}{2h + \gamma|\omega + \Omega|} \biggr|_{\Omega \to +0} \nonumber \\
&\simeq& \frac{2 \gamma k_{\rm B} T h}{\mu(\mu+2h)}, 
\end{eqnarray}
where the quantum fluctuation terms were neglected in obtaining the second line. The corresponding expression in 3D case is 
\begin{equation}
\xi_0 R_q \sigma^{(3)}_{{\rm AL}, xx} \simeq \frac{\gamma k_{\rm B} T}{2 \sqrt{\mu^{(3)}}}. 
\end{equation}
Equation (\ref{rmass2}) is used for the renormalized mass $\mu$ in 2D case, while eq.(\ref{rmass3}) is used for the renormalized mass $\mu^{(3)}$ in 3D case. 

Below, examples of the normalized resistivity curves obtained numerically will be discussed in a manner of comparing results in 2D case with those in 3D case, where the normalized resistivity is given by $R/R_N = \sigma_N/\sigma_{\rm tot}$. First of all, the normalized resistivity vs magnetic field curves taken at various fixed temperatures are examined. We note here that, throughout this work, the field dependences of $R/R_N$ at a fixed temperature are presented for convenience of numerical computations rather than its temperature dependences at a fixed field. Typical curves in 2D case obtained under the fixed values of the fluctuation strength $b_G^{(2)}=0.01$ and the disorder strength $\Delta^{(2)}=10 \Delta_\Phi = 0.002$ are shown in Fig.8. There, since $\theta(t) \simeq 0.1$ even at the lowest temperature $t = 0.318$, the curves in Fig.8 can be regarded as belonging to the category of moderately clean systems. Over most of the field range with nonzero resistance, the normalized resistivity just smoothly decreases with decreasing field. However, close to the end of the smoothly vanishing resistance, a sign of a slight drop of the resistance is seen at each temperature. 

\begin{figure}
\includegraphics{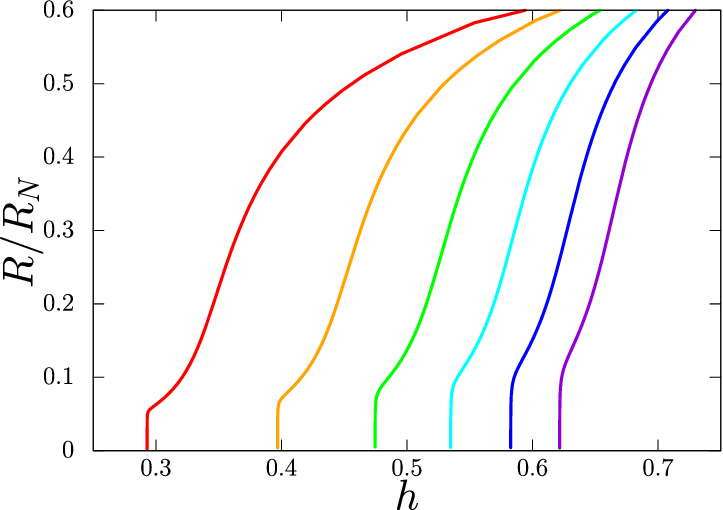}
\caption{Normalized resistivity curves in 3D case corresponding to those in Fig.8 taken at $t=0.637$ (red), $0.531$, $0.455$, $0.398$, $0.354$, and $0.318$ (violet) from left to right. The values $\sqrt{b_G^{(3)}}=0.01$ and $\Delta^{(3)}=10 \Delta_\Phi=0.002$ are used here. }
\label{fig.9}
\end{figure}

Here, we will show that this position of the faint sudden drop seen on each curve in Fig.8 essentially corresponds to the vortex lattice melting field $h_m$ in 2D case in clean limit obtained in the LLL approximation. According to Refs.23 and 24, $h_m$ in clean limit satisfies 
\begin{equation}
1 - t - h_m = c_2^{-1} \sqrt{h_m \, t \, b_G^{(2)}} 
\label{melting2}
\end{equation}
with $c_2 \simeq 0.0988$. By assuming $\Delta^{(2)}$ to be negligibly small and substituting the values of $b_G^{(2)}$ and $t$ into eq.(\ref{melting2}), the obtained value $h_m \simeq 0.10$ for $t=0.637$ coincides with the field at which the red (left) curve reduces to zero in Fig.8. Since, with decreasing $t$, the $\theta$-value increases and the normalized resistivity value is slightly lowered from that expected in clean limit, the $h_m$ value at lower $t$-values becomes slightly lower than the field at which the corresponding resistivity curve reduces to zero. Therefore, it is understood that, at least at weak enough disorder, the 2D VG transition {\it in the Gaussian approximation} occurs at the corresponding vortex lattice melting transition point. Although, of course, the 2D VG transition is believed not to occur in a sophisticated approach beyond the Gaussian approximation \cite{FFH,Moorevg}, this result in the 2D Gaussian approximation plays a valuable role in understanding the corresponding result in 3D case. Hereafter, the field at which the resistivity at weak enough disorder shows a sudden drop will be called $h_{g}$. 

Figure 9 presents the normalized resistivity vs field curves in 3D case at the same set of temperatures obtained using the comparable values of the material parameters $\sqrt{b_G^{(3)}} = 0.01$ and $\Delta^{(3)}=0.002$ with those in Fig.8. Since the effective disorder strength $\theta^{(3)}$ increases with decreasing $t$, the right (violet) curve corresponds to a case with stronger disorder. In contrast to the curves in 2D case, each of the resistance curves in Fig.9 shows a clear and nearly discontinuous drop at a field $h_{g}$ of the type seen in experiments \cite{Nishizaki,Kwok}. As in 2D case, $h_{g}$ will be compared with the melting field $h_m$ expected in clean limit. Evaluation of $h_m$ has been performed based on the thermodynamic data \cite{Schilling} in optimally-doped cuprates and on the LLL fluctuation theory. According to Refs.21 and 22, the melting field $h_m$ in clean limit is given by 
\begin{equation}
1 - t - h_m = c_3^{-1} (h_m \, t )^{2/3} \, (b_G^{(3)})^{1/3} 
\label{melting3}
\end{equation}
with $c_3 \simeq 0.27$. By assuming the disorder strength $\Delta^{(3)}$ to be negligibly small, we find that the $h_m$-value obtained by choosing the value $c_3=0.22$ in eq.(\ref{melting3}) coincides with the $h_{g}$-value of the curve at the highest temperature $t=0.637$ in Fig.9. This implies that the SC transition field in Fig.9 is slightly lower than the melting field expected based on eq.(\ref{melting3}), or equivalently that the fluctuation effect in 3D case is slightly overestimated in our calculation. As already suggested in sec.4, this is a consistent result with our approximation used for the fully-renormalized vertex and the self-energy in 3D case. Therefore, we can expect \cite{FFH,RI96} that, in 3D systems with weak enough disorder, the freezing from the vortex liquid to the vortex solid triggers a sudden vanishing of the resistivity, i.e., the VG ordering. 

\begin{figure}
\includegraphics{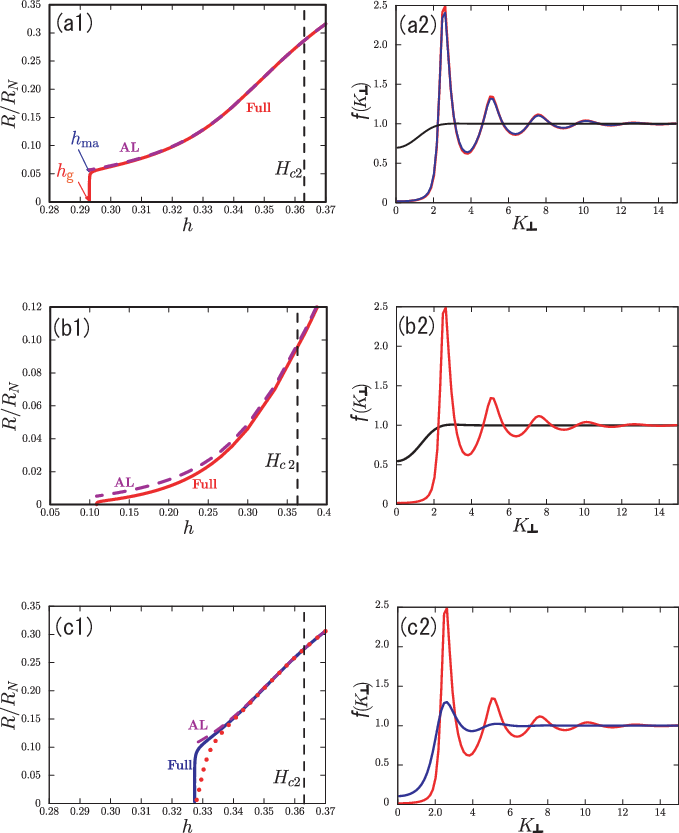}
\caption{(a1) Full: 3D normalized resistivity (red solid) curve at $t=0.637$ in Fig.9. AL: Corresponding (blue dashed) curve obtained by deleting the $\sigma_{{\rm vg},xx}$ term from $\sigma_{\rm tot}$. The dashed vertical line denotes the position of the $H_{c2}$-line, and $h=H/H_{c2}(0)=\xi_0^2/r_H^2$. (a2) Data of the structure factor $f(K_\perp)$ (see eq.(\ref{vcorigin})) taken at $H_{c2}$ (black), at the SC transition field $h_{g}$ (red), and at $h_{ma}$ (blue) slightly above $h_{g}$ indicated on the Full curve of (a1). Here, $K_\perp = |{\bf k}|/\sqrt{h}$ denotes the magnitude of the component perpendicular to the field of the wavevector measured in the unit of $r_H^{-1} = \sqrt{2 \pi H/\phi_0}$. 
(b1) Full: 2D normalized resistivity (red solid) curve at $t=0.637$ in Fig.8. AL: Corresponding (blue dashed) curve obtained by deleting the two $\sigma_{{\rm vg},xx}$ terms from $\sigma_{\rm tot}$. (b2) Data of $f(K_\perp)$ at $H_{c2}$ (black) and at the field $h_{g} = 0.103$ (red) corresponding to the 2D VG transition field {\it in the Gaussian approximation}. 
(c1) Full: 3D normalized resistivity (blue solid) curve at $t=0.637$ in the case with the larger disorder strength $\Delta^{(3)}=0.016$ and the same value of $b_G^{(3)}$ as in (a1).  AL: Corresponding resistivity (blue dashed) obtained by deleting the $\sigma_{{\rm vg},xx}$ term from $\sigma_{\rm tot}$. Explanation on the red dotted curve is given in the text. (c2) Data of $f(K_\perp)$ at the SC transition field $h_{g} = 0.328$ (blue) for the Full curve in (c1). This depressed structure factor implies that the sharp vanishing of the resistivity in (c1) is {\it not} due to the growth of the positional correlation of the vortices. Further details are explained in the text. For comparison, the red curve shown in the figure (a2) is also presented here. }
\label{fig.10}
\end{figure}

In Fig.10 (a), the normalized 3D resistivity curve at the high temperature $t=0.637$ in Fig.9 is compared with structure factor data taken close to $h_{g}$ and at $h_{c2}=H_{c2}(t)/H_{c2}(0)$. The already-mentioned fact that the field $h_g$ at which the resistivity suddenly vanishes coincides with the melting field in clean limit implies that the $t=0.637$ case in Fig.9 is a situation quite close to the clean limit. Therefore, the sudden vanishing of the resistivity in the figure (a1) is a consequence of the growth of the correlation on the vortex positions, as argued previously in Ref.16. It is found in (a2) that, in this case with weak enough disorder, the Bragg peaks just at $h_{g}$ and at a slightly higher field $h_{ma}$ than $h_{g}$ are almost the same as each other. It suggests that there is another origin of the sudden drop of the resistivity other than the growth of the positional correlation. Judging from the fact that the melting field $h_m$ yields the LLL scaling \cite{Thouless,IOT89,Dorsey,Welp}, eq.(\ref{melting3}), on the SC fluctuation in magnetic fields, we argue that, after all, the nature of the SC fluctuation below $H_{c2}$ plays essential roles, together with the growth of the correlation on the vortex position, in determining the vanishing behavior of the resistance at the SC transition point $h_{g}$. In fact, as seen in the figures (a1) and (b1), there are differences in the vortex flow behavior of the resistance in the fluctuation-induced vortex liquid between the 2D and 3D cases. The feature in the figure (b1) that an, if any, sharp drop of the resistance in 2D case merely appears at the end of the vanishing resistivity curve is in contrast to the clear sudden drop of the 3D normalized resistivity in the figures (a1) and (c1). This dimensionality dependence of the vanishing behavior at $h_g$ of the resistivity is a consequence of the difference in the resistivity curve in the vortex liquid, i.e., in the nature of the SC fluctuation properties, between the 2D and 3D cases. 

When compared with resistivity data in cuprates \cite{Nishizaki,IOT91,Kwok}, one might feel that the normalized values of the resistance $R/R_N$ in the vortex liquid regime in Figs.9 and 10 (a1) are too small compared with those in the typical resistance curves. In relation to this, we note that the value $1/(R_q \xi_0)$ we have assumed here for the normal part $\sigma_N$ of the conductivity is too small compared with the typical values of $\sigma_N$ seen in cuprates. To understand the consistency between the result in the figure (a1) and experimental data in clean samples of cuprates, we point out that the jump value of the resistance in (a1) at $h=h_g$ is comparable with the corresponding ones seen in cuprates. In fact, the resistivity curve in (a1) indicates that the SC part of the conductivity, $\sigma_{\rm tot} - \sigma_N$, at $h=h_g$ is about seven times larger than the corresponding value at $h_{c2}=H_{c2}/H_{c2}(0)$. One can verify that this ratio is almost the same as the corresponding value estimated from experimental data on YBCO \cite{Nishizaki,IOT91,Kwok}. 

As can be seen by comparing the figures (a1), (c1), and (c2) with one another, an increase of the disorder leads to an increase of $h_g$ and a shrinkage of the Bragg peak near $h_g$. Accompanying this, the normalized resistivity vs field curve changed from the concave vortex flow behavior like in (a1) to a nearly straight one. The reasonable reduction of the Bragg peak due to the enhanced disorder would imply that the manner of including the disorder effect in the parquet approximation for the vertex function $\Gamma$ is satisfactory. On the other hand, the nearly sharp vanishing of the resistance at $h_g$ in the figure (c1) is a feature disagreeing with experimental data. In fact, it is empirically known that the resistance in the cases with strong disorder, where, as seen in (c2), the vortex positional correlation remains short-ranged, vanishes continuously, i.e., accompanied by a vanishing slope $d R/dT$ \cite{FFH,DHF}. 
One origin of this discrepancy seems to consist in the use of eq.(\ref{appdynG}) for the dynamical SC fluctuation propagator even in dirty cases. As a consequence of the use of eq.(\ref{appdynG}), the dynamical critical exponent $z_{\rm vg}$ for the 3D VG transition is two, and the resistivity in 3D case inevitably vanishes like $\sim (h - h_g)^{1/2}$, i.e., with a divergent slope of $dR/dh$, irrespective of the magnitude $\Delta^{(3)}$ of the sample disorder. However, it is usually expected that, for a continuous VG transition in strongly disordered cases, the vanishing behavior of the resistance is characterized by a dynamical critical exponent $z_{\rm vg}$ larger than four \cite{FFH}. In the Hartree approximation \cite{DHF,II02} neglecting the vertex correction due to the vortex positional ordering to the random potential, the relation $z_{\rm vg}=4$ follows, and a qualitatively reasonable vanishing behavior of the resistivity in a strongly disordered case on approaching the VG transition from above is obtained. At present, it is unclear to us how the results in the Hartree approximation can be approached by starting from the present approach including the effect of the vortex positional ordering on the random potential. For a reference, another resistivity (red dotted) curve is added in the figure (c1). This dotted curve is obtained by assuming $z_{\rm vg}=5$ and replacing the dimensionless quantity $(1 - c_0^{(3)})^{-1/2}$ in the expressions (\ref{xiperp}) and (\ref{xipara}) of the VG correlation lengths $\xi_{vg, \perp}$ and $\xi_{vg, \parallel}$ by $(1 - c_0^{(3)})^{-2}$. Another origin of the discrepancy in the strong disorder case seems to consist in our assumption of weak disorder in forming the VG correlation function in Fig.5:  As the disorder is increased, the irreducible vertex shown in Fig.5 (a) should be formed by not a single disorder line but multiple ones. If this treatment in Fig.5 is to be improved, the resistivity value of the solid curve (indicated as Full) in the figure (c1) should be lowered further in the vortex liquid regime. In any case, if a reasonable interpolation between the clean and dirty cases becomes possible, the disorder dependence of the resistivity curves of the type found in Ref.38 and consequently, the resistive behavior suggesting the presence of the vortex slush regime \cite{Worthington,Nishizaki} below the melting temperature in clean limit could be explained theoretically. 


\section{Summary and Discussion}

In the present paper, the resistive behavior in type II superconductors with weak enough disorder under a magnetic field was studied for both 2D and 3D systems. Empirically, a nearly discontinuous (sharp) vanishing of the resistivity upon cooling is often observed in bulk clean materials and is usually interpreted as an evidence of the first order freezing transition to a vortex solid \cite{Nishizaki,Kwok}, while the resistivity in 2D case decreases smoothly even in relatively clean materials \cite{Saito}. Further, no theory explaining the presence of the vortex slush region found in moderately clean 3D materials \cite{Worthington,Nishizaki} was available. To try to resolve these issues, we have studied the vortex-glass (VG) ordering in type II superconductors with weak enough disorder and have shown that the presence in 3D case of a sharp vanishing of the resistivity and its absence in 2D case are consequences of an interplay of the proximity to the vortex solidification and the nature of the SC fluctuation in the vortex liquid regime. To the best of our knowledge, this is the first theoretical study of realizing the nearly discontinuous vanishing of the resistivity in moderately clean 3D case through detailed calculations. 

On the other hand, the resistive behavior characteristic of the vortex slush region, i.e., a sharp drop of the resistivity signaling the vortex solidification followed by its continuous vanishing, cannot be explained within the present approach because of our use of the Gaussian approximation for the VG fluctuation. Our theory predicts that, in clean enough 3D systems, the vortex lattice melting temperature in clean limit becomes the onset of the VG fluctuation \cite{RI96} upon sweeping the temperature. However, the use of the Gaussian VG fluctuation with the exponent $0.5$ of the conductivity does not lead to a tail of the resistivity following its sharp vanishing. Such a tail becomes more visible in a dirtier system with a wider VG critical region. Thus, a visible vortex slush region would be created by renormalizing the VG fluctuation. Further, in 2D case, such renormalized VG fluctuations should delete any vanishing of the resistivity \cite{FFH} and would clarify the validity of the picture \cite{RI96,Nunchot23} that the vortex lattice melting would not be reflected in the resistivity in relatively clean 2D systems \cite{Shahar}. 

It should be stressed that developing a theory on the vortex phase diagram in a high field region is important in relation to the recent observation of new fluctuation regions associated with the paramagnetic pair-breaking \cite{Nunchot22} and the quantum fluctuation \cite{Nigel} in iron-based superconductors \cite{Kasahara,Hardy}, because the approaches constructed based on the phase-only model \cite{Natter,Blatter,GleD}, developed in relation to the study on the vortex states in cuprates, are not applicable to the vortex states in such a high field regime. In fact, a sharp vanishing of the resistance accompanied by a broad vortex liquid regime has appeared even in the high field regime in 
FeSe \cite{Kasahara}.

\begin{acknowledgement}
The present work was supported by a Grant-in-Aid for Scientific Research [No.21K03468] from the Japan Society for the Promotion of Science. 
\end{acknowledgement}

\appendix*\section{}

To clarify symmetries in the relations of eq.(36), we note that the following relations,  
\begin{equation}
{\hat \Gamma}_{abcd}({\bf k}) = \Gamma_{abdc}({\bf k}), 
\end{equation}
and 
\begin{equation}
{\hat C}({\bf k}) = {\hat A}({\bf k}) {\hat B}({\bf k})
\end{equation}
for $C({\bf k})=(A \, * \, B)({\bf k})$, are satisfied together with the last equation of (\ref{parquetcomplex}). 
These two relations follow from their definitions. By applying them to the relations of eq.(36), we find 
\begin{eqnarray}
\Pi_{2,abcd}({\bf k}) &=& {\hat \Pi}_{3,abdc}({\bf k}), \nonumber \\
\Lambda_{2,abcd}({\bf k}) &=& {\hat \Lambda}_{3,abdc}({\bf k}). 
\end{eqnarray}

It is straightforward to find 
\begin{eqnarray}
\Pi_{1,abcd}({\bf k}) &=& {\hat \Pi}_{1,abdc}({\bf k}), \nonumber \\
\Lambda_{1,abcd}({\bf k}) &=& {\hat \Lambda}_{1,abdc}({\bf k}) 
\end{eqnarray}
in a similar manner. 

\section{}
\renewcommand\theequation{B.\arabic{equation}}

To evaluate the expressions $J_1^{(2)}$ and $J_2^{(2)}$ appearing in the conductivities, the expression on a pair of the propagators 
\begin{eqnarray}
p_{\mu_1,\mu_2} &=& \biggl(-\frac{\partial}{\partial \Omega} \biggr) \sum_\omega {\cal G}_1(\omega) {\cal G}_2(\omega+\Omega) \biggl|_{\Omega \to +0} 
= \frac{\gamma}{2} \sum_\omega \biggl( {\cal G}_1(\omega) {\cal G}_2(\omega)[{\cal G}_1(\omega) + {\cal G}_2(\omega)] \nonumber \\ 
&-& [({\cal G}_1(\omega))^2 + ({\cal G}_2(\omega))^2] \frac{{\cal G}_1(0) {\cal G}_2(0)}{{\cal G}_1(0) + {\cal G}_2(0)} \biggr)
\label{AL}
\end{eqnarray}
becomes useful, where ${\cal G}_n(\omega)=1/(\mu_n+\gamma|\omega|)$. Using eq.(\ref{AL}), one obtains 
\begin{equation}
J_1^{(2)} \simeq [G_{\rm vg}({\bf k}; 0,0)]^2 \lim_{T \to \infty} p_{\mu,\mu} = \frac{\gamma}{2} [{\cal G}(0)]^3 [G_{\rm vg}({\bf k}; 0,0)]^2, 
\end{equation}
where the quantum fluctuations with nonzero $\omega$ were neglected by taking the limit $T \to \infty$. 
This result on $J_1^{(2)}$ can also be found by expanding $G_{\rm vg}({\bf k}; 0, 0)$ in a power series in $X({\bf k}) [{\cal G}(0)]^2$ and using the relation 
\begin{eqnarray}
p^{(mn)}_{\mu,\mu}|_{cl} &\equiv& \lim_{T \to \infty} \biggl(-\frac{\partial}{\partial \Omega} \biggr) \sum_{\omega} {\cal G}^m(\omega) {\cal G}^n(\omega+\Omega) \biggl|_{\Omega \to +0} = \gamma (m+1) {\cal G}^{m+n+3}(0) \nonumber \\
&\times& \biggl[ 1 - \frac{1}{2^{m+n+1}} \biggl( \frac{(2m+1) (m+n)!}{n! (m+1)!} - \frac{m-n}{m+1} \sum_{r=0}^{m-1} \frac{1}{2^r} \frac{(n+r)!}{n! r!} \biggr) \biggr]
\label{appb}
\end{eqnarray}
as the main term of an expression following from resumming up 
the power series. 

In a similar way, $J_2^{(2)}$ is evaluated using eq.(\ref{appb}).


\begin{thebibliography}{9}
\bibitem{com} Any theoretical work based on the LLL approach, which is essentially a high field approximation, is inapplicable to systems with a freezing to a vortex solid occurring in exceptionally low fields such as the Bi-compounds of cuprates. Therefore, in comparing theoretical results with experimental data, we will focus on the materials in which a melting or glass transition occurs in a relatively high field range.
\end{thebibliography}
\end{document}